\documentclass[prx,a4paper,portrait,aps,twocolumn,superscriptaddress,10pt]{revtex4-1}

\usepackage{amsmath,amsthm,amssymb,bbm,amsfonts,graphics,xcolor,times,xfrac,booktabs,mathtools,xr,subfigure,bbm,verbatim,appendix,placeins,comment,mathrsfs}
\usepackage[unicode=true,bookmarks=true,bookmarksnumbered=false,bookmarksopen=false,breaklinks=false,pdfborder={0 0 1}, backref=false,colorlinks=true]{hyperref}
\renewcommand*{\url}[1]{\href{#1}{#1}}
\usepackage{orcidlink}
\usepackage[shortlabels]{enumitem}

\makeatletter
\theoremstyle{plain}
\newtheorem{thm}{\protect\theoremname}
\theoremstyle{plain}
\newtheorem{lem}{\protect\lemmaname}
\theoremstyle{plain}

\theoremstyle{remark}
\newtheorem*{rem*}{\protect\remarkname}
\theoremstyle{plain}

\theoremstyle{plain}

\theoremstyle{definition}
\newtheorem{defn}{\protect\definitionname}
\theoremstyle{plain}

\theoremstyle{plain}
\newtheorem*{thm*}{\protect\theoremname}
\theoremstyle{plain}
\newtheorem*{lem*}{\protect\lemmaname}
 
\providecommand{\propositionname}{Proposition}
\providecommand{\theoremname}{Theorem}
\providecommand{\lemmaname}{Lemma}
\providecommand{\remarkname}{Remark}
\providecommand{\conjecturename}{Conjecture}
\providecommand{\definitionname}{Definition}
\providecommand{\corollaryname}{Corollary}
\providecommand{\observationname}{Observation}
\allowdisplaybreaks

\def\bra#1{\langle{#1}\vert}
\def\ket#1{\vert{#1}\rangle}

\def\ketbra#1#2{\vert{#1}\rangle\!\langle{#2}\vert}

\newcommand{\ot}{\otimes}

\def\tr#1{\mbox{tr}\left[{#1}\right]}

\hypersetup{
	bookmarksnumbered,
	pdfstartview={FitH},
	citecolor={darkgreen},
	linkcolor={darkred},
	urlcolor={darkblue},
	pdfpagemode={UseOutlines}}
\definecolor{darkgreen}{RGB}{50,190,50}
\definecolor{darkblue}{RGB}{0,0,190}
\definecolor{darkred}{RGB}{238,0,0}
\definecolor{quantum}{RGB}{83,37,127}
\definecolor{quantumlight}{RGB}{169,146,191}
\definecolor{darkorange}{RGB}{255,100,0}
\usepackage{soul}

\usepackage{pifont}

\usepackage{array}

\newcommand{\nl}{\ensuremath{\hspace*{-0.5pt}}}
\newcommand{\nr}{\ensuremath{\hspace*{0.5pt}}}

\newcommand{\subtiny}[3]{\ensuremath{_{\hspace{#1 pt}\protect\raisebox{#2 pt}{\tiny{$ #3$}}}}}
\newcommand{\suptiny}[3]{\ensuremath{^{\hspace{#1 pt}\protect\raisebox{#2 pt}{\tiny{$ #3$}}}}}

\makeatletter
\theoremstyle{plain}

\providecommand{\propositionname}{Proposition}
\providecommand{\theoremname}{Theorem}
\providecommand{\lemmaname}{Lemma}
\providecommand{\remarkname}{Remark}
\providecommand{\conjecturename}{Conjecture}
\providecommand{\definitionname}{Definition}
\providecommand{\corollaryname}{Corollary}
\providecommand{\observationname}{Observation}
\allowdisplaybreaks

\def\bra#1{\langle{#1}\vert}
\def\ket#1{\vert{#1}\rangle}

\def\BraVert{e.g.,roup\,\mid\,\bgroup}

\def\tr#1{\mbox{tr}\left[{#1}\right]}
\newcommand{\ptr}[2]{\mbox{tr}_{\raisebox{-1pt}{\tiny{$#1$}}}\left[ #2 \right]}

\DeclareMathOperator{\diag}{diag}

\makeatletter
\newsavebox\myboxA
\newsavebox\myboxB
\newlength\mylenA
\newcommand*\xoverline[2][0.75]{%
    \sbox{\myboxA}{$\m@th#2$}%
    \setbox\myboxB\null
    \ht\myboxB=\ht\myboxA%
    \dp\myboxB=\dp\myboxA%
    \wd\myboxB=#1\wd\myboxA
    \sbox\myboxB{$\m@th\overline{\copy\myboxB}$}
    \setlength\mylenA{\the\wd\myboxA}
    \addtolength\mylenA{-\the\wd\myboxB}%
    \ifdim\wd\myboxB<\wd\myboxA%
       \rlap{\hskip 0.5\mylenA\usebox\myboxB}{\usebox\myboxA}%
    \else
        \hskip -0.5\mylenA\rlap{\usebox\myboxA}{\hskip 0.5\mylenA\usebox\myboxB}%
    \fi}
\makeatother

\begin{document}

\title{Efficiently Cooling Quantum Systems with Finite Resources: \\ Insights from Thermodynamic Geometry}

\author{Philip Taranto\,\orcidlink{0000-0002-4247-3901}}
\email{philip.taranto@manchester.ac.uk} 
\thanks{P. T. and P. L.-B. contributed equally.}
\affiliation{Department of Physics \& Astronomy, University of Manchester, Manchester M13 9PL, United Kingdom}
\affiliation{Department of Physics, Graduate School of Science, The University of Tokyo, 7-3-1 Hongo, Bunkyo City, Tokyo 113-0033, Japan}

\author{Patryk Lipka-Bartosik\,\orcidlink{0000-0001-5254-1245}}
\affiliation{Department of Applied Physics, University of Geneva, 1211 Geneva, Switzerland}
\affiliation{Institute of Theoretical Physics, Faculty of Physics, Astronomy and Applied Computer Science, Jagiellonian University, 30-348 Kraków, Poland}

\author{Nayeli A. Rodr{\' i}guez-Briones\,\orcidlink{0000-0002-3945-518X}}
\affiliation{Atominstitut, Technische Universit{\"a}t Wien, Stadionallee 2, 1020 Vienna, Austria}
\affiliation{Miller Institute for Basic Research in Science, University of California Berkeley, CA 94720, USA}

\author{\mbox{Mart{\' i} Perarnau-Llobet}\,\orcidlink{0000-0002-4658-0632}}
\affiliation{F{\' i}sica Te{\` o}rica: Informaci{\' o} i Fen{\` o}mens Qu{\` a}ntics, Departament de F{\' i}sica, Universitat Aut{\` o}noma de Barcelona, 08193 Bellatera (Barcelona), Spain}
\affiliation{Department of Applied Physics, University of Geneva, 1211 Geneva, Switzerland}

\author{Nicolai Friis\,\orcidlink{0000-0003-1950-8640}}
\affiliation{Atominstitut, Technische Universit{\"a}t Wien, Stadionallee 2, 1020 Vienna, Austria}
\affiliation{Institute for Quantum Optics and Quantum Information - IQOQI Vienna, Austrian Academy of Sciences, Boltzmanngasse 3, 1090 Vienna, Austria}

\author{Marcus Huber\,\orcidlink{0000-0003-1985-4623}}
\email{marcus.huber@tuwien.ac.at}
\affiliation{Atominstitut, Technische Universit{\"a}t Wien, Stadionallee 2, 1020 Vienna, Austria}
\affiliation{Institute for Quantum Optics and Quantum Information - IQOQI Vienna, Austrian Academy of Sciences, Boltzmanngasse 3, 1090 Vienna, Austria}

\author{Pharnam Bakhshinezhad\,\orcidlink{0000-0002-0088-0672}}
\email{pharnam.bakhshinezhad@tuwien.ac.at} 
\affiliation{Atominstitut, Technische Universit{\"a}t Wien, Stadionallee 2, 1020 Vienna, Austria}

\begin{abstract}
Landauer's limit on heat dissipation during information erasure is critical as devices shrink, requiring optimal pure-state preparation to minimise errors. However, Nernst's third law states this demands infinite resources in energy, time, or control complexity. We address the challenge of cooling quantum systems with \emph{finite} resources. Using Markovian collision models, we explore resource trade-offs and present efficient cooling protocols (that are optimal for qubits) for coherent and incoherent control. Leveraging thermodynamic length, we derive bounds on heat dissipation for swap-based strategies and discuss the limitations of preparing pure states efficiently.
\end{abstract}
\maketitle


\textit{Introduction.---}One of the most essential tasks in quantum science is preparing pure quantum states, equivalent to cooling physical systems or erasing information. This is a critical prerequisite for quantum computation, where the output state from a calculation must be erased before it can be reused as an input for the next~\cite{Bennett_1982}. Failure to create sufficiently pure states leads to computational errors and reduces the accuracy of timekeeping~\cite{Xuereb_2023,SchwarzhansLockErkerFriisHuber2021} and measurement~\cite{Guryanova_2020}. Without adequate purity, possibly due to limited resources or control, the frequency of gate and measurement errors increases, potentially relegating any `quantum advantage' to mere conjecture.

In this sense, thermodynamics links the degree of \emph{control} over a system with one's \emph{capacity} to perform useful tasks. Landauer established that a minimum amount of heat must be dissipated when erasing information encoded in \emph{any} physical system, formalising a connection between physics and information~\cite{Landauer_1961}. This limit applies to classical and quantum theory, gaining prominence as computing elements are miniaturised and become more susceptible to heat-induced errors. Efforts to saturate the Landauer bound involve engineering quasistatic interactions between information-bearing systems and controllable machines. However, determining the necessary conditions for Landauer-cost erasure has been impeded by inequivalent assumptions across experimental platforms. 

A breakthrough by Reeb and Wolf reformulated the Landauer limit in the context of quantum information, providing platform-agnostic insights~\cite{Reeb_2014}. They demonstrated the need for an infinitely large energy gap in an infinite-dimensional machine to achieve perfect Landauer-cost erasure. Yet, infinite resources are not practically accessible, leading to the challenge of optimising cooling with finite resources~\cite{meier2023,munson2024complexityconstrained}. When resources are limited, factors such as the energy-level structure of the cooling machines and the complexity of their interactions with the target system influence the achievable final purity and associated energy cost, leading to a three-way trade-off among \emph{energy}, \emph{time}, and \emph{control complexity}~\cite{Taranto_2023}. 

Here, we explore the relationship between energy cost and time in the finite-resource regime. This setting is difficult to analyse in general, as cooling performance depends on a complex interplay of microscopic details. To make progress, we focus on cooling procedures implemented via a Markovian collision model~\cite{Rau_1963,Ziman_2002,Scarani_2002,Ziman_2005,Ciccarello_2017,Taranto_2020}, where the target system is cooled through a sequence of unitary interactions with uncorrelated thermal machines. We connect these models to continuous trajectories in the state space and use the geometric technique of thermodynamic length~\cite{weinhold1975metric,Salamon1983,Crooks_2007,SivakCrooks,Scandi2019thermodynamiclength,Abiuso2020} to bound the cooling performance. For a simple but insightful collision model based on swap operations, we derive the associated thermodynamic metric and optimal cooling protocols for the case of qubit systems. Our work contributes to the understanding of resource limitations for the important task of preparing pure states. This represents a first step into connecting the framework of thermodynamic geometry with (Markovian) collision models, establishing methods that may prove valuable for analysing cooling procedures in more complicated settings. 


\emph{Framework.---}We begin with the setting.

\begin{figure}[t]
\centering
\hspace*{-4mm}\includegraphics[width=0.75\linewidth]{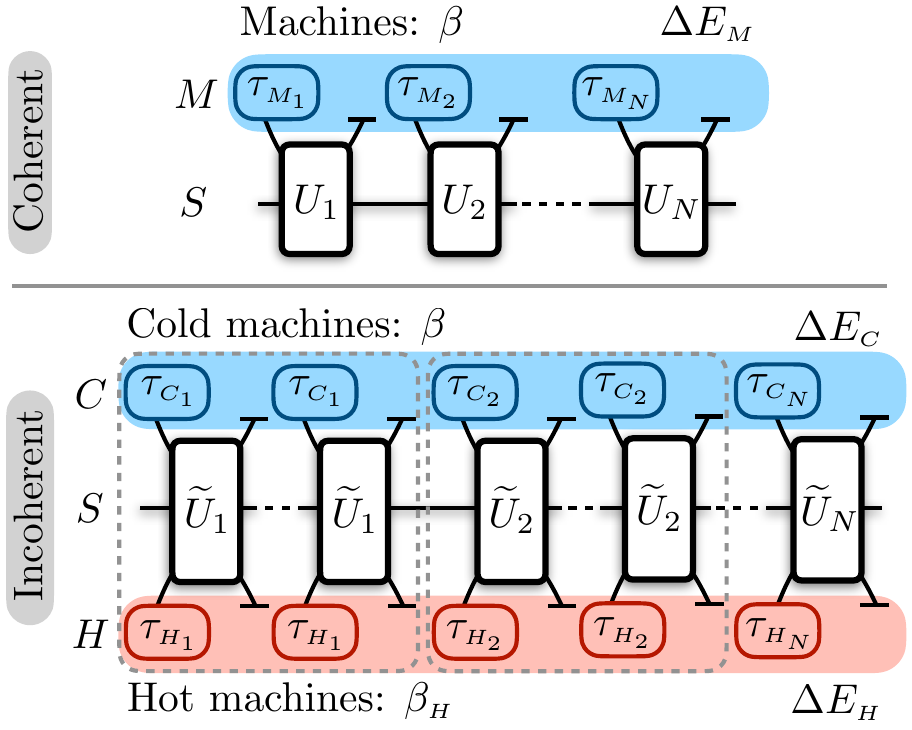}
\vspace*{-2mm}
\caption{\textit{Framework.}
A system $S$ is cooled via sequential interactions with a system $M$ comprising $N$ machines, using a Markovian collision model. In the coherent-control setting (top), arbitrary unitaries $U_i$ act between $S$ and initially thermal machines $\tau\subtiny{0}{0}{M_i}(\beta,H\subtiny{0}{0}{M_i})$. In the incoherent-control setting (bottom), $M$ splits into hot $(H)$ and cold ($C$) systems at inverse temperatures $\beta\subtiny{0}{0}{H}$ and $\beta$, respectively. Here, the unitaries are restricted to be energy conserving, i.e., $[\widetilde{U}_i,H\subtiny{0}{0}{S}+H\subtiny{0}{0}{C_i}+H\subtiny{0}{0}{H_i}]=0$. Moreover, we allow repeated interactions between the target and copies of hot and cold machines, which we call a \emph{stage} (grey outline). The energy cost is the change in energy of the appropriate machines, $\Delta E\subtiny{0}{0}{X}$ for $X \in \{ M, C, H\}$\@. \vspace{-1em}} \label{fig::framework} 
\end{figure}

\noindent\textbf{Setting. }We consider cooling a target quantum system~$S$ via unitary interactions with another system~$M$\@, composed of~$N$ subsystems $\{M_1,\hdots,M_N\}$ called \emph{machines}. We describe the procedure by a Markovian \emph{collision model}~\cite{Rau_1963,Ziman_2002,Scarani_2002,Ziman_2005,Ciccarello_2017,Taranto_2020}. Here, the target unitarily interacts with a fresh machine at each time, reflecting the property of memorylessness and rapid machine rethermalisation between control operations (see Fig.~\ref{fig::framework}). 

All systems $X\!\in\!\{ S, M_1,\hdots,M_N\}$ have an associated Hilbert space $\mathcal{H}\subtiny{0}{0}{X}$, on which states $\varrho\subtiny{0}{0}{X}$ are represented as positive semidefinite, unit-trace operators. Each system has a Hamiltonian whose spectral decomposition fixes its energy structure, $H\subtiny{-1}{0}{X}\!:=\!\sum_{i=0}^{d\subtiny{0}{0}{X}-1} E\subtiny{0}{0}{X}\suptiny{1}{0}{(i)} \ket{i}\!\bra{i}$\@. We consider finite-dimensional systems $d\subtiny{0}{0}{X}\! :=\! \mathrm{dim}(\mathcal{H}\subtiny{0}{0}{X}) \! <\! \infty$, and assume that $E\subtiny{0}{0}{X}\suptiny{1}{0}{(i\!+\!1)}\! \geq\! E\subtiny{0}{0}{X}\suptiny{1}{0}{(i)}$, with $E\subtiny{0}{0}{X}\suptiny{1}{0}{(0)} := 0$\@. With respect to any Hamiltonian $H$, the thermal state at inverse temperature $\beta := (k\subtiny{0}{0}{\mathrm{B}} T)^{-1}$ is $\tau(\beta,H) := \mathcal{Z}^{-1}(\beta) \exp{(-\beta H)}$, where $\mathcal{Z}(\beta) := \tr{\exp{(-\beta H)}}$ is the partition function; when unambiguous, we will write $\tau(\beta)$. The thermal state uniquely maximises entropy $S(\varrho) := -\tr{\varrho \log{\varrho}\nr}$ for fixed average energy $E(\varrho) := \tr{H \varrho\nr}$\@, providing a suitable initial machine state for \emph{cooling schemes} (formalised below).\\[-3mm]

\noindent\textbf{Boundary Conditions. }We investigate procedures that transition a target system from an initial state $\varrho\subtiny{0}{0}{S}$ to a final one $\varrho'\subtiny{0}{0}{S}$ while concurrently transforming the collection of machines from an initial thermal state $\tau\subtiny{0}{0}{M}(\beta)$ to a final state $\varrho\subtiny{0}{0}{M}^\prime$\@. This transformation occurs via the global evolution described by:
\begin{align}\label{eq::unitarycooling}
    \varrho'\subtiny{0}{0}{SM} = U [\varrho\subtiny{0}{0}{S}\otimes \tau\subtiny{0}{0}{M}(\beta)] U^\dagger.
\end{align}
The goal of cooling is to manipulate the target so that the majority of its populations are transferred to the lowest energy eigenstates. Any meaningful notion of cooling can be captured by a majorisation relation (see SM Sec.~I~\cite{SM}); for simplicity, we focus on processes that take the target from a Gibbs state characterised by an initial $\beta$ to a final one with $\beta_{\mathrm{f}} := \lambda \beta$, where $\lambda > 1$. This fixes the boundary conditions. \\[-3mm]


\noindent\textbf{Structural and Control Resources.} Cooling performance is influenced by several factors, including the dimensions $d\subtiny{0}{0}{X}$ and Hamiltonians $H\subtiny{-1}{0}{X}$ of all systems, the interaction range $k$ (the number of systems involved in each interaction), the number of machines $N$, and the dissipated heat $\Delta E\subtiny{0}{0}{M} := \tr{H\subtiny{0}{0}{M} (\varrho\subtiny{0}{0}{M}^\prime - \varrho\subtiny{0}{0}{M})}$, which establishes a lower bound for the energetic cost of any implementation. We distinguish \emph{structural} resources---such as $d\subtiny{0}{0}{X}$ and $H\subtiny{-1}{0}{X}$, which are fixed independently of the procedure---from \emph{control} resources linked to the protocol, such as the interaction range $k$, total time duration (the number of timesteps $N$ for fixed control complexity $k$), and the dissipated heat $\Delta E\subtiny{0}{0}{M}$\@. \\[-3mm] 

\noindent\textbf{Type of Control.} We consider two extremal control paradigms: \emph{coherent} and \emph{incoherent} (see Fig.~\ref{fig::framework})~\cite{Clivaz_2019L,Clivaz_2019E,Taranto_2023}. Coherent control allows a work source to implement any system-machine unitary, enabling arbitrary transformations as described in Eq.~\eqref{eq::unitarycooling}. In contrast, incoherent control employs energy-conserving unitaries between the target and machines at different temperatures to drive all heat and entropy flows. Coherent control represents the highest level of control in a thermodynamic setting, while incoherent control assumes less control, requiring only the switching on and off of interaction Hamiltonians. The settings of heat-bath algorithmic cooling~\cite{Baugh_2005,Schulman_2005,Raeisi_2015,RodriguezBriones_2016,Park_2016,Alhambra_2019,Lin_2024} and of autonomous cooling~\cite{Linden_2010,Levy_2012,Brunner_2012,Mitchison_2016,Maslennikov_2019,Manikandan_2020,Guzman_2023}
are contained within these paradigms, respectively. \\[-3mm] 

\noindent\textbf{Cooling Schemes.} We now define the concept of a cooling scheme, encompassing all aforementioned dependencies.\vspace{-0.6mm}
\begin{defn}
    A \emph{cooling scheme}
    is defined by the tuple: $(\mathscr{B},\mathscr{S},\mathscr{C},\mathscr{T})$. Here, $\mathscr{B}$ denotes the \emph{boundary conditions} of the problem, namely, the initial and final temperature of the target system. The \emph{structural resources} $\mathscr{S}$ include $\beta$, $H\subtiny{-1}{0}{X}$, and $d\subtiny{0}{0}{X}$, i.e., the initial temperature, Hamiltonians, and dimensions of all systems. The \emph{control resources} $\mathscr{C}$ encompass the total number of machines $N$, the interaction range $k$, and the energy cost $\Delta E\subtiny{0}{0}{M}$\@. Finally, the \emph{type} $\mathscr{T}$ indicates whether the procedure operates with the coherent or incoherent control. 
\end{defn}\vspace{-0.6mm}

Notably, Nernst's third law of thermodynamics and Landauer's bound exemplify instances where particular resource configurations preclude achievability. Nernst's law states that infinite resources are required to prepare a pure state~\cite{Nernst_1906,Ticozzi_2014,Masanes_2017}; in our context, this implies an infinitely large energy gap in the machine is necessary~\cite{Reeb_2014,Clivaz_2019L,Taranto_2023}. Similarly, Landauer's bound establishes that the entropy of the target cannot be reduced by $\widetilde{\Delta} S := S(\varrho\subtiny{0}{0}{S}) - S(\varrho\subtiny{0}{0}{S}^{\prime})$ via interactions with a thermal machine without dissipating at least $\beta \Delta E\subtiny{0}{0}{M}$~\cite{Landauer_1961,Bennett_1982,Esposito_2011,Reeb_2014}. Delineating the boundary of attainable cooling procedures for different resource configurations represents a significant open problem~\cite{Taranto_2023,munson2024complexityconstrained}. Here, we focus on achievable schemes, optimising over finite resources to develop effective cooling procedures. Specifically, we seek the optimal energy-level structure of machines $\{ H_{M_i} \}_i$ and interactions $\{ U_{i} \}_i$ to minimise the energy cost of a cooling scheme characterised by fixed control complexity $k$ and finite duration, represented by the number of steps $N$. To achieve this objective, we leverage the geometric concept of thermodynamic length~\cite{weinhold1975metric,Crooks_2007,SivakCrooks,Abiuso2020}.


\textit{Thermodynamic Geometry.---}The notion of thermodynamic length enables the characterisation of path-dependent thermodynamic quantities, such as (dissipated) work, via a geometric approach~\cite{Nulton1985,Crooks_2007,Abiuso2020}. To define this concept, consider a path in Hamiltonian space $H(t)$ parameterised by $t \in [ 0, 1 ]$. The \emph{thermodynamic length} associated to such a path is given by
\begin{align}\label{eq:thermodynamic-length}
    \mathcal{L} := \int_{0}^{1} \sqrt{\mathrm{cov}_{t}\bigl(\beta\dot{H}(t),\beta\dot{H}(t)\bigr)}\,\text{d}t.
\end{align}
where $\dot{H}(t) := \partial_t H(t)$ and 
\begin{align}
    \text{cov}_{t}(A, B) := \tr{\mathcal{J}_{t}(A)B} - \tr{\tau(t) A} \tr{\tau(t) B},
\end{align}
with $\mathcal{J}_{t}(A) := \int_{0}^1 \tau(t)^{1-x} A \tau(t)^x \, \textup{d} x$ and $\tau(t) := \tau(\beta, H(t))$. The length squared $\mathcal{L}^2$ is related to the dissipated heat or excess work when slowly driving $H(t)$ whilst in contact with a bath at inverse temperature $\beta$~\cite{Salamon1983,Crooks_2007,SivakCrooks,Scandi2019thermodynamiclength}. The minimal length connecting two endpoints $H(t_0)$ and $H(t_1)$ minimises the dissipation along a path in Hamiltonian space and is found by solving the geodesic equations; for commuting Hamiltonians, an analytic solution to Eq.~\eqref{eq:thermodynamic-length} is known~\cite{Jencova2004}.

The notion of thermodynamic length is typically employed in slowly driven systems ~\cite{SivakCrooks,Scandi2019thermodynamiclength,Abiuso2020,BrandnerSaito,MehboudiMiller,GengChenSunDong,FrimDeWeeseMichael} or discrete-time processes~\cite{Nulton1985,Scandi2020}. We will show that this useful concept can also be applied to the setting of Markovian collision models in the limit of a large number of machines $N$. The key observation here is that slowly changing the system's Hamiltonian can be approximated by a sequence of simple swap interactions in which the system interacts with thermal machine subsystems with carefully chosen Hamiltonians. This approach enables to characterise the back-action of the process in the machine or bath, in contrast to previous works, e.g., Refs.~\cite{Nulton1985,SivakCrooks,Scandi2020}. This connection allows us to develop new insights to our question of interest: Given the ability to apply $N$ unitary interactions (of fixed complexity $k$) between system and machines, what is the optimal machine energy-level distribution for cooling the target system? 


\textit{Coherent Control.---}In the coherent-control setting, given infinite resources, the Landauer bound sets the ultimate limit on cooling. We first focus on the role of finite structural complexity in said scenarios. We strive to identify the structural complexity that minimises the energy cost when the system is cooled via a sequence of $N\! <\! \infty$ bipartite ($k\! =\! 2$) interactions. 

\begin{thm}\label{thm::thermodynamiclengthcoherent}
In the coherent-control setting, given a qudit system with Hamiltonian $H\subtiny{0}{0}{S}$, the minimum energy cost of cooling from a thermal state $\tau\subtiny{0}{0}{S}(\beta, \, H\subtiny{0}{0}{S})$ to $\tau\subtiny{0}{0}{S}( \lambda \beta, H\subtiny{0}{0}{S})$ using swaps between the system and a fresh qudit machine with arbitrary Hamiltonian at each of the $N$ steps, is given by 
\begin{align}\label{eq::coherentthermodynamiclength}
    \beta \Delta E\subtiny{0}{0}{M} = \widetilde{\Delta} S\subtiny{0}{0}{S} + \frac{1}{2 N} (\mathcal{L}^{*})^{2} + \mathcal{O}(N^{-2}),
\end{align}
where $\mathcal{L}^*$ is the minimal thermodynamic length~\cite{Jencova2004,Abiuso2020}: 
\begin{align}\label{eq::coherentthermolengthoptimal}
    \mathcal{L}^* &= 2 \arccos \left(\mathrm{tr}\left[\sqrt{\tau\subtiny{0}{0}{S}(\beta, \, H\subtiny{0}{0}{S})}\sqrt{\tau\subtiny{0}{0}{S}( \lambda \beta, H\subtiny{0}{0}{S})}\right]\right) \notag \\
    &= 2 \arccos \left(\frac{\mathcal{Z}(\beta(1+\lambda)/2)}{\sqrt{\mathcal{Z}(\beta) \mathcal{Z}(\lambda \beta)}}\right).
\end{align}
\end{thm}
\noindent Above, $\beta \Delta E\subtiny{0}{0}{M}$ is the (scaled) energy cost of cooling, $\widetilde{\Delta} S\subtiny{0}{0}{S}$ is the Landauer bound (the minimum energy cost), and the remaining terms represent the additional finite-resource penalty. As $N$ increases, this additional cost decreases like $1/N$, approaching the Landauer bound. The thermodynamic length $\mathcal{L}^*$ quantifies how far apart the initial and final states are in a thermodynamic sense; the greater this distance, the more energy above Landauer cost is required for finite $N$.\\

\noindent\textbf{Sketch of proof}.\ 
The proof, fully detailed in Supplemental Material \textbf{(SM)} Sec.~II~\cite{SM}, is based on the equality form of Landauer's bound~\cite{Reeb_2014}:
\begin{align}
    \beta\Delta E\subtiny{0}{0}{M} = \widetilde{\Delta} S\subtiny{0}{0}{S} + I(S\!:\!M)_{\varrho\subtiny{0}{0}{SM}^\prime} + D(\varrho\subtiny{0}{0}{M}^\prime\nr\|\nr\tau\subtiny{0}{0}{M}(\beta)),
\end{align}
which holds for any transformation described by Eq.~\eqref{eq::unitarycooling} such that the entropy of the target changes from $S(\varrho\subtiny{0}{0}{S})$ to $S(\varrho\subtiny{0}{0}{S}^\prime)=S(\varrho\subtiny{0}{0}{S})-\widetilde{\Delta} S\subtiny{0}{0}{S}$. Here, $I(X\!:\!Y)_{\varrho\subtiny{0}{0}{XY}}:=S(\varrho\subtiny{0}{0}{X})+S(\varrho\subtiny{0}{0}{Y})-S(\varrho\subtiny{0}{0}{XY})$ is the quantum mutual information and $D(\varrho\subtiny{0}{0}{X}\nr\|\nr\varrho\subtiny{0}{0}{Y}\!) := \tr{\varrho\subtiny{0}{0}{X} (\log{\varrho\subtiny{0}{0}{X}}- \log{\varrho\subtiny{0}{0}{Y}}\!)}$ is the quantum relative entropy. This equality breaks down the total energy cost into three terms: the fundamental Landauer bound $\widetilde{\Delta} S\subtiny{0}{0}{S}$, correlations built up between system and machine $I(S\!:\!M)_{\varrho\subtiny{0}{0}{SM}^\prime}$, and a term measuring how far the machine's final state deviates from thermal equilibrium $D(\varrho\subtiny{0}{0}{M}^\prime\nr|\nr\tau\subtiny{0}{0}{M}(\beta))$. The proof then proceeds in two steps: First, the bipartite interactions are chosen to be swaps between the qudit system and each of a sequence of qudit machines with increasing energy gaps, such that no correlations are built up between $S$ and $M$ as the system is cooled, i.e., $I(S\!:\!M)_{\varrho_{SM}^\prime} = 0$. Then, the relative-entropy term is minimised; for the sequence of swap operations considered, the relative entropy has the tight lower bound $\tfrac{1}{2N} (\mathcal{L}^*)^2 + \mathcal{O}(N^{-2})$, and we thus assert the claim. \hfill\qed\\

\begin{figure}
    \centering
    \includegraphics[width=0.9\linewidth]{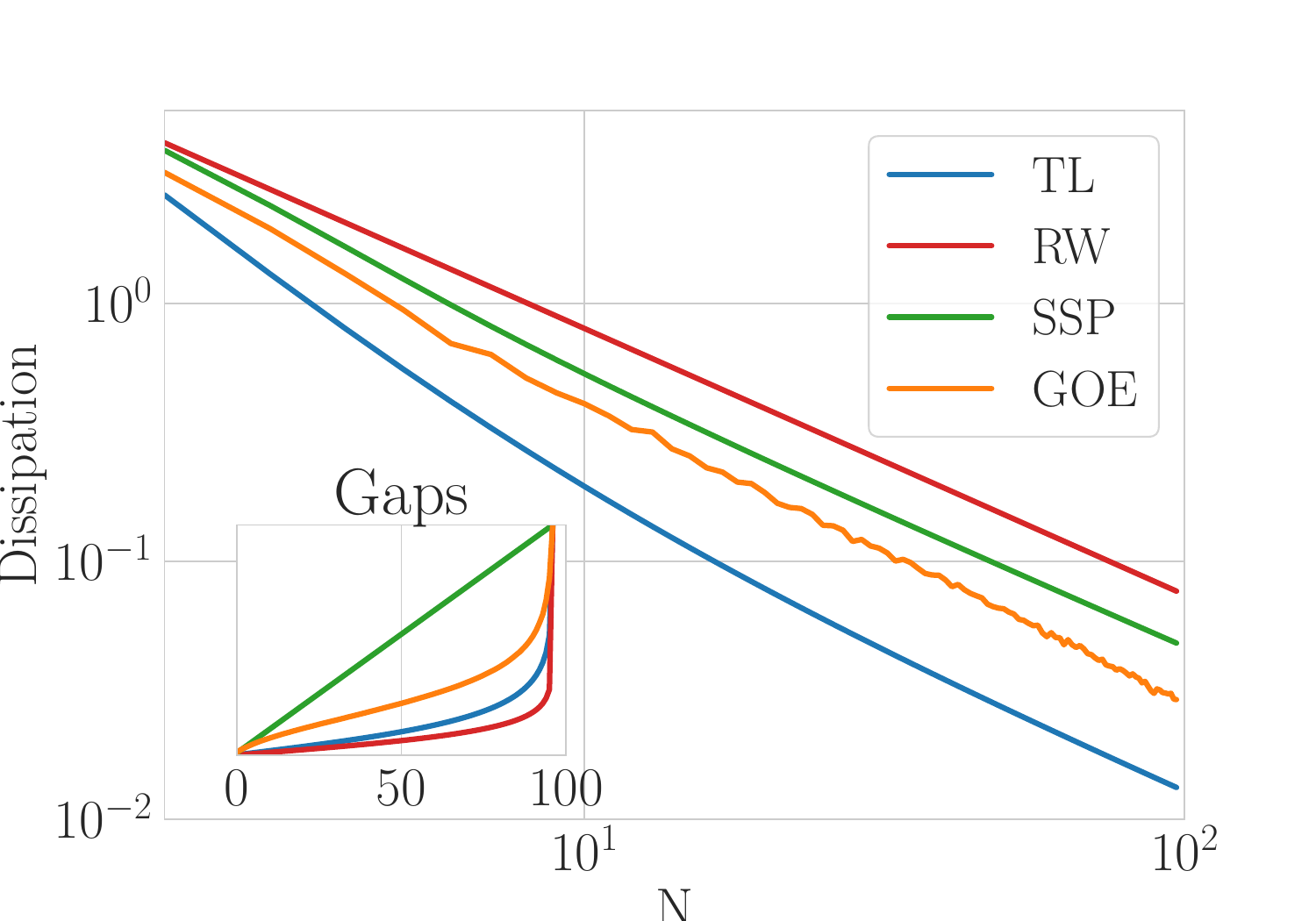}
    \caption{\textit{Cooling with Coherent Control.} Heat dissipation when coherently cooling a qubit target with energy gap $E\subtiny{0}{0}{S} = 1$ from $\beta = 10^{-8}$ to $\beta_{\mathrm{f}} = 10$ (i.e., $\lambda = 10^9$) via sequential bipartite interactions with $N$ machine qubits. We compare four procedures that depend on the energy-level structure $\{H_{M_i}\}_i$, whose gaps are depicted in the inset for $N = 100$. TL is the optimal protocol deduced from the thermodynamic length (see Theorem~\ref{thm::thermodynamiclengthcoherent}); RW corresponds to the protocol from Ref.~\cite{Reeb_2014} where machine energy gaps change linearly; SSP corresponds to the protocol from Ref.~\cite{skrzypczyk2014work} where the machine qubits' excited state populations change linearly. GOE is a protocol where the machine gaps correspond to energy level spacings drawn from the Gaussian orthogonal ensemble (see SM Sec. III~\cite{SM}); interestingly, this procedure also outperforms the RW and SSP protocols. \vspace{-2em}}
    \label{fig:coh_protocols}
\end{figure}

To summarise, having fixed the parameters $d\subtiny{0}{0}{X} < \infty$, $N < \infty$, and $k=2$, but not the structural complexity, i.e., the machine Hamiltonians $\{ H\subtiny{0}{0}{M_i} \}_i$, 
of a coherent-control cooling scheme, 
we have optimised the remaining control-resource parameter, namely the energy cost $\Delta E\subtiny{0}{0}{M}$\@. In the case of qubit target and machines, we show that such swaps constitute the optimal interaction (see SM Sec.~II A~\cite{SM}) and derive the Hamiltonian that saturates Eq.~\eqref{eq::coherentthermodynamiclength} (see SM Sec.~II B~\cite{SM}), thereby providing the optimal cooling scheme with respect to heat dissipation in the case of qubits. In Fig.~\ref{fig:coh_protocols}, we compare this optimal protocol with other known coherent cooling schemes to demonstrate its effectiveness. Although the optimal energy structure in the case of swaps for higher dimensions is given by Eq.~\eqref{eq::coherentthermolengthoptimal}, determining the optimal operation in general remains an open problem. Indeed, when considering higher-dimensional machines (including those with interacting subsystems), further cooling advantages can be achieved~\cite{Rolandi2023,Lipka_Perarnau_2024} (see also SM Sec.~II C~\cite{SM}).

Some comments regarding optimality are in order. First, we are assuming that the cooling procedure is Markovian, i.e., that the machines are completely refreshed between steps of controlled evolution. In this setting, creating correlations costs energy~\cite{Huber_2015,BruschiPerarnauLlobetFriisHovhannisyanHuber2015,Bakhshinezhad_2019}, which implies that the optimal scheme must minimise the correlations built up between system and machines~\cite{Afsary_2020, Lipka_2024}. However, in a non-Markovian setting, correlations could potentially be used in later steps to lower the energy cost or improve performance~\cite{Rodriguez-briones_2017PRL,Rodriguez-briones_2017,Taranto_2020}. Second, there is a non-zero energy cost for creating coherences~\cite{Misra_2016}. Since we assume an initial state that is diagonal in the energy eigenbasis, this implies that the optimal cooling scheme must only permute populations of energy eigenstates (leading to a final system state that commutes with the initial one \cite{Abiuso2020}), but for general initial states this need not be the case. Nonetheless, since we evaluate the cooling performance in terms of the heat dissipated by the machine and neglect any system-local energy cost (i.e., that associated to local basis changes), our results apply to arbitrary initial and final system states.

\textit{Incoherent Control.---}We now consider the same question in the incoherent-control setting. In contrast to the coherent-control setting, here all heat and entropy flows occur solely by coupling the target system to a hot ($H$) and cold ($C$) machine, leading to an energy-conserving transformation overall. Beginning with $\varrho\subtiny{0}{0}{SC{\nl}H} = \tau\subtiny{0}{0}{S}(\beta) \otimes \tau\subtiny{0}{0}{C}(\beta) \otimes \tau\subtiny{0}{0}{H}(\beta\subtiny{0}{0}{H})$, where $\beta\subtiny{0}{0}{H} < \beta$, the considered evolution leads to the output state $\varrho\subtiny{0}{0}{SC{\nl}H}^\prime = \widetilde{U} \varrho\subtiny{0}{0}{SC{\nl}H} \widetilde{U}^\dagger$, where $[\widetilde{U},H\subtiny{0}{0}{S}+H\subtiny{0}{0}{C}+H\subtiny{0}{0}{H}]=0$ encodes energy conservation, i.e., that the total energy remains constant throughout. Note that in the incoherent-control scenario, no energy-conserving unitary with another single thermal machine can lead to cooling the target~\cite{Clivaz_2019E}; hence, interactions between at least three systems must be considered.

In this setting, the Landauer bound is unattainable; instead, the ultimate limit is given by the Carnot-Landauer bound~\cite{Taranto_2023}
\begin{align}\label{eq:landauerincoherent2}
\Delta F\subtiny{0}{0}{S}^{\beta} + \eta \Delta E\subtiny{0}{0}{H}  &\leq 0,
\end{align}
which follows from the equality form
\begin{align}
\Delta F\subtiny{0}{0}{S}^{\beta} + \eta\, \Delta E\subtiny{0}{0}{H}  &=\,-\beta^{-1}[\Delta S\subtiny{0}{0}{S}\,+\,\Delta  S\subtiny{0}{0}{C}\,+\,\Delta  S\subtiny{0}{0}{H}\,\label{eq:landauerincoherent1}\\*[1mm]
&\ \ +\,D(\varrho\subtiny{0}{0}{C}^\prime\nr\|\nr\tau\subtiny{0}{0}{C}(\beta))\,+\, D(\varrho\subtiny{0}{0}{H}^\prime\nr\|\nr\tau\subtiny{0}{0}{H}(\beta\subtiny{0}{0}{H}))]. \nonumber
\end{align} 
Here, we have introduced the free energy $F\subtiny{0}{0}{X}^{\beta}(\varrho) := \tr{H\subtiny{-1}{0}{X} \varrho\nr} - \beta^{-1} S(\varrho)$ and the Carnot efficiency $\eta := 1-\beta\subtiny{0}{0}{H}/\beta \in [0,1]$\@.

In a similar vein to the coherent-control scenario, we wish to bound the right-hand side of Eq.~(\ref{eq:landauerincoherent1}) for any finite-resource implementation, and ideally identify a protocol that saturates this bound. However, a number of problems immediately arise in the incoherent-control scenario, since one is restricted to the orbit of energy-conserving unitaries, i.e., $\widetilde{U}$ such that $[\widetilde{U},H\subtiny{0}{0}{S}+H\subtiny{0}{0}{C}+H\subtiny{0}{0}{H}]=0$\@. This constraint implies that the relative-entropy terms cannot be bounded simply by the thermodynamic length, which was possible in the coherent-control setting because the full swap led to a straightforward expression in terms of a sequence of relative-entropy terms applied to the chain of machines. Here, such a swap is prohibited by energy conservation. We now present an attainable energy bound for finite-resource cooling with incoherent control, which is generally optimal for qubits, and optimal for qudits within the considered class of interactions, in analogy to Theorem~\ref{thm::thermodynamiclengthcoherent} in the coherent-control setting.

\begin{thm}\label{thm::thermodynamiclengthincoherent}
In the incoherent-control setting, given a qudit system with Hamiltonian $H\subtiny{0}{0}{S}$, the minimum energy cost for cooling from a thermal state $\tau\subtiny{0}{0}{S}(\beta, \, H\subtiny{0}{0}{S})$ to  $\tau\subtiny{0}{0}{S}( \lambda  \beta, H\subtiny{0}{0}{S})$, by using particular tripartite (energy-conserving) interactions between the system and two fresh qudit machines at inverse temperature $\beta$ (cold) and $\beta\subtiny{0}{0}{H} \leq \beta$ (hot), respectively, with arbitrary Hamiltonians in the limit of infinite steps but with $N$$<\infty$ distinct energy gaps, is given by 
\begin{gather}\label{eq::incoherentqubitthm}  
\Delta F\subtiny{0}{0}{S}^\beta\,+\eta \,\Delta E\subtiny{0}{0}{H} = \,-\frac{1}{2N\beta} (\mathcal{L}^{*})^2 +\mathcal{O}(N^{-2}).
\end{gather}
\end{thm}

\noindent This result shows that, like in the coherent case, the additional energy cost above the fundamental limit (here the Carnot-Landauer bound) scales as $1/N$ in the finite-resource setting. \\

\noindent\textbf{Sketch of proof}.\ The proof, presented in SM Sec.~IV~\cite{SM}, is fundamentally different to its coherent-control counterpart. In the constructive direction, we propose a cooling scheme comprising interactions that exchange populations amongst levels $\ket{i, i+1, i}\subtiny{0}{0}{SCH} \leftrightarrow \ket{i+1,i,i+1}\subtiny{0}{0}{SCH}$. The energy-conserving nature allows us to calculate the energy cost per population transfer, which is related to the relative entropy between the initial and final states of the virtual-qubit subspaces of the hot-and-cold machine that permit cooling. We finally bound this quantity by the thermodynamic length.\hfill\qed\\

\noindent For qudits, it is not clear that the form of energy-conserving interactions considered here are optimal; nonetheless, within this family, we present a cooling scheme that attains the energy cost of Eq.~\eqref{eq::incoherentqubitthm} and saturates the Carnot-Landauer bound in the limit of infinitely many distinct energy gaps, i.e., diverging control complexity. In the special case of cooling a qubit target with (hot and cold) qubit machines, we show that the cycle $\ket{010}\subtiny{0}{0}{SC{\nl}H} \leftrightarrow \ket{101}\subtiny{0}{0}{SC{\nl}H}$ is indeed optimal. This is because for any fixed set of energy gaps, the family of energy-conserving unitaries on three qubits that permit cooling without creating coherences or correlations must be of this form, and thus we can cover the entire orbit of unitaries in question. Such operations can be considered as a \emph{virtual} swap between the target and the \emph{virtual} qubit subspace of the machine spanned by $\{ \ket{01}\subtiny{0}{0}{CH}, \ket{10}\subtiny{0}{0}{CH}\}$\@. In general, since such a subspace has norm strictly less than one, each such virtual swap will lead to the system qubit being at strictly higher temperature than the virtual qubit. However, in the limit of infinitely many repetitions within a single stage, the temperature of the system's qubit subspace of interest converges to the virtual temperature of the machine-qubit subspace~\cite{Clivaz_2019L,Clivaz_2019E}. As we are interested in finite resources, we assume that one performs a finite but sufficiently large number of virtual swaps so that the error is within specified tolerances. The relative entropy term that governs the finite-cooling behaviour here (and which leads to the thermodynamic-length term) concerns the initial and final thermal states of the machine at the corresponding \emph{virtual temperature} defined by the qubit subspace in question. Implementing the protocol that swaps the target successively with appropriately chosen virtual qubits of the machine in each stage minimises the thermodynamic length and therefore provides the optimal incoherent cooling procedure. 


\textit{Role of Correlations.---}The constraint of energy conservation distinguishes the paradigms of coherent and incoherent control. In the latter setting, the virtual subspaces spanned by the hot-and-cold machines influence the performance of a cooling scheme, rather than the state of the machine per se. This suggests that \emph{correlations} play a dominant role in the incoherent-control setting; we now formalise this intuition. 

\begin{thm}\label{thm::incoherentcorrelations}
    For any incoherent-control cooling scheme, the sum of free energy differences $\Delta F\subtiny{0}{0}{X}^\beta$ (w.r.t. inverse temperature $\beta$) is bounded by the sum of generated correlations $\Delta I_{\alpha}$, 
    \begin{align}
    \sum\subtiny{0}{0}{X\!\in\!\{\!S\!,\!C\!,\!H\!\}} \Delta F\subtiny{0}{0}{X}^\beta \leq -\tfrac{2}{3} \beta^{-1} \sum_{\alpha} \Delta I_{\alpha},
    \end{align}
    where $\alpha \in\{SC,SH,CH\}$, and $I\subtiny{0}{0}{XY} := I(X\!:\!Y)_{\varrho\subtiny{0}{0}{XY}}$ is the quantum mutual information.
\end{thm}

\noindent A proof is given in SM Sec.~V~\cite{SM}. This bound has interesting implications. For instance, a priori, the only claim that one can deduce regarding the free energy change of the hot machine is $\Delta F\subtiny{0}{0}{H}^\beta \leq 0$, which follows from both the system and cold machine beginning in thermal states at inverse temperature $\beta$. However, using the relation $\beta \Delta F\subtiny{0}{0}{X}^\beta = D(\varrho\subtiny{0}{0}{X}^\prime \nr\|\nr \tau\subtiny{0}{0}{X}(\beta))$ for $X \in \{ S, C\}$, we can derive the tighter bound
\begin{align}
 \beta \Delta F\subtiny{0}{0}{H}^\beta \leq - \tfrac{2}{3}\! \sum_{\alpha} \! \Delta I_\alpha - D(\varrho\subtiny{0}{0}{S}^\prime \nr\|\nr \tau\subtiny{0}{0}{S}(\beta)) - D(\varrho\subtiny{0}{0}{C}^\prime \nr\|\nr \tau\subtiny{0}{0}{C}(\beta)) \leq 0,
\end{align}
where the second inequality follows from the non-negativity of both the mutual information and the relative entropy. 


\textit{Discussion.---}Efficient cooling of quantum systems in practice necessitates the optimisation of machines and interactions over complicated resource constraints. Here, we have made several contributions to this problem, focussing primarily on qubit systems while developing methods that may prove valuable in more general settings. Firstly, we formalised the concept of a cooling scheme using a universal definition that captures all relevant dependencies, permitting fair comparison among different procedures. Secondly, for the case of fixed control complexity, we demonstrated simple protocols that asymptotically saturate the ultimate bounds and dissipate minimal heat in the regime of many (but finite) machines, establishing their optimality specifically for qubit systems under both coherent and incoherent control. Our main technical contribution bounds the heat dissipated by the machine in a cooling process in terms of the geometric concept of thermodynamic length. We make this connection by modelling cooling processes as Markovian collision models, linking prominent methods used in quantum thermodynamics and information theory. Finally, we analysed the role of correlations in the incoherent-control setting, deriving a bound on free-energy differences in terms of correlations. 

A key direction for future work is the extension of our analysis to higher-dimensional systems, where the optimality of our protocols remains an open question. While our framework provides a foundation for such investigations, the optimisation problem becomes considerably more complex beyond the qubit setting, potentially requiring different mathematical techniques and approaches. Other important directions include extending our analysis beyond Markovian collision models and developing protocols that saturate the correlation bounds in the incoherent-control setting. Yet, like other higher-dimensional problems at the intersection of thermodynamics and information theory (e.g., regarding symmetrically thermalising unitaries~\cite{Bakhshinezhad_2019}), general solutions and optimality proofs may be difficult to obtain due to the large parameter spaces involved. In light of this observation, more pragmatic platform-specific approaches may be called for, and we hence envisage future attempts to address such questions to be tailored to more particular (experimental) setups. \vspace{-1em}


\begin{acknowledgments}
We thank Fabien Clivaz, Morteza Rafiee, and Ralph Silva for insightful discussions. P. T. acknowledges funding from the Japan Society for the Promotion of Science (JSPS) Postdoctoral Fellowships for Research in Japan and the IBM-UTokyo Laboratory. P.L.-B. acknowledges the Swiss National Science Foundation (SNSF) through project 192244 and the Polish National Science Centre through project Sonata 2023/37351/D/ST2/02309. M. P.-L. acknowledges the Swiss National Science Foundation for financial support through Ambizione grant PZ00P2-186067 and also acknowledges funding from the Spanish Agencia Estatal de Investigacion through the grant  ``Ram{\'o}n y Cajal RYC2022-036958-I''. N. A. R.-B. acknowledges the support from the Miller Institute for Basic Research in Science, at the University of California, Berkeley. N. F. acknowledges financial support from the Austrian Science Fund (FWF) through the stand-alone project P 36478-N funded by the European Union – NextGenerationEU, as well as by the Austrian Federal Ministry of Education, Science and Research via the Austrian Research Promotion Agency (FFG) through the flagship project  FO999897481 funded by the European Union – NextGenerationEU. This publication was made possible through the support of Grant 62423 from the John Templeton Foundation. The opinions expressed in this publication are those of the author(s) and do not necessarily reflect the views of the John Templeton Foundation. M. H., N. A. R.-B., and P. B. would like to acknowledge funding from the European Research Council (Consolidator grant `Cocoquest' 101043705). M. H. further acknowledges funding by FQXi (FQXi- IAF19-03-S2, within the project ``Fueling quantum field machines with information'') as well as the European flagship on quantum technologies (`ASPECTS' consortium 101080167). 
\end{acknowledgments}

\def\bibsection{\section*{References}} 
\bibliographystyle{apsrev4-1fixed_with_article_titles_full_names_new}

%

\newpage
\onecolumngrid
\appendix
\section*{Appendices}
\setcounter{section}{0}

\section{Notions of cooling}\label{app::notionofcooling} 

To begin with, note that cooling a physical system can have several inequivalent meanings. For equilibrium processes, it could mean reducing the temperature of a thermal state; in non-equilibrium settings, it could mean increasing the ground-state population or purity of the target, or decreasing its entropy or average energy. The strongest notion of cooling derives from the preorder on states induced by \emph{majorisation}: We say that a state $\varrho$ is colder than $\sigma$ iff $\varrho \succ \sigma$. Since all meaningful notions of temperature are Schur-convex/concave functions of the vector of non-decreasing energy eigenvalues, any other such quantifier would agree that $\varrho$ is colder than $\sigma$~\cite{Clivaz2020Thesis}. Although all qualitative results presented here hold true in this strong sense.

\section{Optimal and efficient cooling with coherent control (Proof of Theorem~\ref{thm::thermodynamiclengthcoherent})}\label{app::thermodynamiclengthcoherent} 

\begin{proof}\textbf{(Theorem~\ref{thm::thermodynamiclengthcoherent}).} Let $S$ be a qudit of dimension $d$ with a local Hamiltonian $H\subtiny{0}{0}{S}$ and let $M$ be a system of dimension $d\subtiny{0}{0}{M} = d\suptiny{0}{0}{N}$ with Hamiltonian $H\subtiny{0}{0}{M} = \sum_{n=1}^{N}  \mathbbm{1}\subtiny{0}{0}{d}^{\otimes (n-1)} \otimes H\subtiny{0}{0}{M_n}\otimes \mathbbm{1}\subtiny{0}{0}{d}^{\otimes (N-n)}$ where $H\subtiny{0}{0}{M_n}=\sum_{i=1}^{d}{E}^{(i_n)}\subtiny{0}{0}{M_n} \ket{i_n}\!\bra{i_n}$\@. The joint system consisting of $S$ and $M$ begins in the state $\tau\subtiny{0}{0}{S}(\beta) \otimes \tau\subtiny{0}{0}{M}(\beta)$ and undergoes a unitary process $U$, i.e.,
\begin{align}
    \varrho^\prime\subtiny{0}{0}{SM} = U[\tau\subtiny{0}{0}{S}(\beta) \otimes \tau\subtiny{0}{0}{M}(\beta)]U^{\dagger}.
\end{align}
Our goal is to perform the transformation $\tau\subtiny{0}{0}{S}(\beta) \rightarrow \tau\subtiny{0}{0}{S}(\beta_{\mathrm{f}})$, where $\beta_{\mathrm{f}} > \beta$ (cooling). The energy cost of the transformation specified by $U$ is given by $\Delta E = \Delta E\subtiny{0}{0}{S} + \Delta E\subtiny{0}{0}{M}$ with $\Delta E\subtiny{0}{0}{X} := \tr{H\subtiny{-1}{0}{X}(\varrho^\prime\subtiny{0}{0}{X} - \tau\subtiny{0}{0}{X}(\beta))}$\@. Since $\Delta E\subtiny{0}{0}{S}$ is fixed by the boundary conditions, we focus on quantifying $\Delta E\subtiny{0}{0}{M}$\@. 

To prove the constructive direction of Theorem~\ref{thm::thermodynamiclengthcoherent}, we choose the unitary transformation
\begin{align}
    \qquad U = \mathbb{S}\subtiny{0}{0}{N} \mathbb{S}\subtiny{0}{0}{2} \ldots \mathbb{S}\subtiny{0}{0}{1}, 
    \label{eq:ChoiceU}
\end{align}
where $\mathbb{S}_n$ is a unitary operator swapping the machine subsystem $M_n$ with the target $S$\@. In SM Sec.~\ref{app::qubitoptimality}, we will demonstrate that, for any cooling scheme that is implemented via a Markovian collision model, such a sequence of swap operations is optimal whenever all systems involved are qubits. The above transformation maps $\tau\subtiny{0}{0}{S}(\beta) \ot \tau\subtiny{0}{0}{M}(\beta)$ to $\tau\subtiny{0}{0}{S}(\beta') \ot \varrho^\prime\subtiny{0}{0}{M}$ where 
\begin{align}
    \tau\subtiny{0}{0}{M}(\beta) = \tau\subtiny{0}{0}{M_1}(\beta) \ot \tau\subtiny{0}{0}{M_2}(\beta) \ot \ldots \ot \tau\subtiny{0}{0}{M_N}(\beta), \qquad
    \varrho^\prime\subtiny{0}{0}{M} = \tau\subtiny{0}{0}{M_0}(\beta) \ot \tau\subtiny{0}{0}{M_1}(\beta) \ot \ldots \ot \tau\subtiny{0}{0}{M_{N-1}}(\beta),
\end{align}
and we have defined $\tau\subtiny{0}{0}{M_0}(\beta) := \tau\subtiny{0}{0}{S}(\beta)$. The energy cost of \emph{any} globally unitary cooling scheme with a thermal machine is~\cite{Reeb_2014}
\begin{align}
    \beta\Delta E\subtiny{0}{0}{M} = \widetilde{\Delta} S\subtiny{0}{0}{S} + I(S\!:\!M)_{\varrho^\prime_{SM}} + D(\varrho^\prime\subtiny{0}{0}{M}\nr\|\nr\tau\subtiny{0}{0}{M}(\beta)). 
\end{align}
Due to our specific choice of $U$ we have $I(S\!:\!M)_{\varrho^\prime_{SM}} = 0$ and furthermore $D(\varrho^\prime\subtiny{0}{0}{M}\nr\|\nr\tau\subtiny{0}{0}{M}(\beta)) = \sum_{n=0}^{N-1} D(\tau\subtiny{0}{0}{M_{n}}(\beta)\nr\|\nr\tau\subtiny{0}{0}{M_{n+1}}(\beta))$\@. Before proceeding further, it is useful to introduce $\Delta H\subtiny{0}{0}{M_n} :=  H\subtiny{0}{0}{M_{n+1}} -  H\subtiny{0}{0}{M_n}$ which will be considered small in what follows. Note that we can always do this, as we are free to choose the Hamiltonian structure $\{H_{M_n}\}$. More specifically, we require that $\Delta H\subtiny{0}{0}{M_n}$ is small enough so that the following perturbative expansion is justified \cite{hiai2014introduction}:
\begin{align}
    \label{eq:tau_n}
    \tau\subtiny{0}{0}{M_{n+1}}(\beta) = \tau\subtiny{0}{0}{M_{n}}(\beta) - \beta \mathcal{J}_n[\Delta_n(\Delta H_{M_n})]\Delta_n(\Delta H_{M_n}),
\end{align}
where we have introduced
\begin{align}
      \Delta_{n}(X) &:= X - \tr{\tau\subtiny{0}{0}{M_n}(\beta) X},
\end{align}
and the operator $\mathcal{J}_n[\cdot\nr] := \mathcal{J}_{\tau\subtiny{0}{0}{M_n}(\beta)}[\cdot\nr]$ with $\mathcal{J}_{\varrho}[A] := \int_{0}^1  \varrho^{1-x} A \varrho^{x} \,\text{d}x$\@ . We now observe that for any density matrix $\varrho$ and traceless operator $\theta$ one can perturbatively expand the relative entropy as \cite{hiai2014introduction}
\begin{align}
    \label{eq:3}
    D(\varrho + \epsilon \theta\nr\|\nr\varrho) = \frac{\epsilon^2}{2} \tr{\theta \mathcal{J}_{\varrho}^{-1}(\theta)} + \mathcal{O}(\epsilon^3),
\end{align}
where $\mathcal{J}_{\varrho}^{-1}$ is an operator formally defined as $\mathcal{J}_{\varrho}^{-1}[A] := \int_{0}^{\infty}  (\varrho + x\ \mathbbm{1})^{-1} A (\varrho + x\  \mathbbm{1})^{-1} \,\text{d}x$, such that it satisfies $\mathcal{J}_{\varrho}^{-1} [\mathcal{J}_{\varrho}[A]] = A$\@. By performing the perturbative expansion from Eq.~\eqref{eq:3} on $D(\tau\subtiny{0}{0}{M_{n}}(\beta)\nr\|\nr\tau\subtiny{0}{0}{M_{n+1}}(\beta))$ with $\epsilon = \
\|\Delta H_{M_n}\|_{\text{tr}} = \mathcal{O}(1/N)$ where $\|\cdot\|_{\text{tr}}$ is the trace norm, and invoking Eq.~\eqref{eq:tau_n}, we obtain 
\begin{align}
    D(\tau\subtiny{0}{0}{M_{n}}(\beta)\nr\|\nr\tau\subtiny{0}{0}{M_{n+1}}(\beta)) =  \frac{\beta^2}{2} \tr{ \mathcal{J}_{n} [\Delta_{n}(\Delta H\subtiny{0}{0}{M_n})] \Delta_n(\Delta H\subtiny{0}{0}{M_n})} + \mathcal{O}\left({N^{-3}}\right).
\end{align}
With this, we can now write
\begin{align}
    D(\sigma\subtiny{0}{0}{M}\nr\|\nr\tau\subtiny{0}{0}{M}(\beta)) &= \sum_{n=0}^{N-1} D(\tau\subtiny{0}{0}{M_{n}}(\beta)\nr\|\nr\tau\subtiny{0}{0}{M_{n+1}}(\beta)) + N \mathcal{O}(N^{-3}) \notag \\* &= \frac{\beta^2}{2 N} \sum_{i=0}^{N-1} \frac{1}{N} \tr{\mathcal{J}_{n} [\Delta_{n}(\dot{H}\subtiny{0}{0}{M_n})] \Delta_{n}(\dot{H}\subtiny{0}{0}{M_n})} + \mathcal{O}({N^{-2}}) \notag \\*
    &= \frac{\beta^2}{2 N} \int_{0}^{1} \tr{\mathcal{J}_{t} [\Delta_{t}(\dot{H}_{t})] \Delta_t(\dot{H}_t)} \text{d}t + \mathcal{O}({N^{-2}}) \notag \\*
    & = \frac{\beta^2}{2 N} \int_{0}^{1} \text{cov}_{t}(\dot{H}_t, \dot{H}_t)\,\text{d}t + \mathcal{O}({N^{-2}}) \notag \\*
    &\geq \frac{1}{2 N} \mathcal{L}^2 + \mathcal{O}(N^{-2}),\label{eq::app-thermolengthbound}
\end{align}
where we have defined $\dot{H}\subtiny{0}{0}{M_n}$ via $\Delta H\subtiny{0}{0}{M_n} = \frac{1}{N} \dot{H}\subtiny{0}{0}{M_n} + \mathcal{O}({N^{-2}})$ in the second line. In the third line we used the fact that $N \gg 1$ and introduced a parameter $t$ with $\textup{d}t = 1/N$ which allowed us to replace the summation with an integral up to an error of $\mathcal{O}(N^{-2})$\@. In the fourth line we introduced $\text{cov}_{t}(A, B) := \tr{\mathcal{J}_{t}(A)B} - \tr{\varrho A} \tr{\varrho B}$\@. Finally, in the last line we introduced the thermodynamic length, i.e.,
\begin{align}
\label{eq:thermo_length_def}
    \mathcal{L} := \int_{0}^{1} \sqrt{\text{cov}_{t}(\beta\dot{H}_t, \beta\dot{H}_t)}\,\text{d}t.
\end{align}
The inequality in Eq.~\eqref{eq::app-thermolengthbound} is saturable, i.e., there exists a protocol that achieves equality. To derive this, we first parameterise 
\begin{align}
\label{eq:ham_param}
    H_t = H_0 + \sum_{i=1}^d\xi_i(t) X_i,
\end{align}
where $\{X_i\}_{i=1}^d$ is an operator basis in the $d$-dimensional Hilbert space. The optimal trajectory $\{\xi_i(t)\}_{i=1}^d$ minimising the thermodynamic length $\mathcal{L}$ can be found by solving the Euler-Lagrange equations:
\begin{align}
    \label{eq:app-euler-lagrange}
    \frac{\partial \mathcal{L}}{\partial \xi_i} = \frac{\text{d}}{\text{d} t} \frac{\partial \mathcal{L}}{\dot{\xi}_i} \qquad \text{for}\quad i = 1, \ldots, d. 
\end{align}
Consequently the energy change of the machine $M$ in the optimal protocol is given by 
\begin{align}\label{eq::app-thermodynamiclengthoptequality}
    \beta\Delta E\subtiny{0}{0}{M} = \widetilde{\Delta} S\subtiny{0}{0}{S} +  \frac{1}{2 N} (\mathcal{L}^*)^2 + \mathcal{O}(N^{-2}),
\end{align}
where $\mathcal{L}^*$ is the thermodynamic length computed for the optimal trajectory (geodesics) $\{\xi_i^*(t)\}_{i=1}^d$, i.e., the solution to Eq.~\eqref{eq:app-euler-lagrange}. \end{proof}


\subsection{Optimality of the sequence of swap operations for cooling a qubit target via a Markovian collision model with qubit machines}\label{app::qubitoptimality}

Above we have provided a constructive cooling scheme that results in an energy cost given by Eq.~\eqref{eq::app-thermodynamiclengthoptequality} for arbitrary systems and thermal machines. In the case where the system-machine interactions consist of a sequence of swap gates, the thermodynamic-length trajectory defined as the solution to Eq.~\eqref{eq:app-euler-lagrange} is optimal. To prove optimality in general we must now argue that the sequence of swap operations assumed at the outset in Eq.~\eqref{eq:ChoiceU} is indeed optimal in terms of heat dissipation amongst all possible unitary transformations that achieve a desired amount of cooling. 

Concretely, the question is: \emph{For a given initial and final state of the target system, related by a unitary evolution with a thermal machine state, what is the transformation that achieves the desired transformation whilst dissipating the least heat?} More formally, consider a target system initially in the state $\varrho\subtiny{0}{0}{S}$. The task in any single step of the cooling procedure is to transform it to some final state $\varrho\subtiny{0}{0}{S}^\prime$ according to
\begin{align}\label{eq::optimality-1}
    \varrho\subtiny{0}{0}{S}^\prime = \ptr{M}{U (\varrho\subtiny{0}{0}{S} \otimes \tau\subtiny{0}{0}{M}(\beta,H\subtiny{0}{0}{M})) U^\dagger}.
\end{align}
We aim to do so in such a way as to minimise the dissipated heat
\begin{align}\label{eq::optimality-2}
    \Delta E\subtiny{0}{0}{M} = \tr{ H\subtiny{0}{0}{M} (\varrho\subtiny{0}{0}{M}^\prime - \tau\subtiny{0}{0}{M}(\beta,H\subtiny{0}{0}{M})) },
\end{align}
where
\begin{align}
    \varrho\subtiny{0}{0}{M}^\prime = \ptr{S}{U (\varrho\subtiny{0}{0}{S} \otimes \tau\subtiny{0}{0}{M}(\beta,H\subtiny{0}{0}{M})) U^\dagger}.
\end{align}
We seek the combination of $U$ and $H\subtiny{0}{0}{M}$ that achieves a given transformation $\varrho\subtiny{0}{0}{S} \mapsto \varrho\subtiny{0}{0}{S}^\prime$ according to Eq.~\eqref{eq::optimality-1} and that minimises Eq.~\eqref{eq::optimality-2}. This can be formally cast as the following optimisation problem:
\begin{align}\label{eq::optimisationdensityoperators}
    &\text{given:} \quad\quad &&\varrho\subtiny{0}{0}{S}, \, \varrho\subtiny{0}{0}{S}^\prime, \,\beta\notag \\
    &\text{minimise:} \quad\quad &&\tr{ H\subtiny{0}{0}{M} \{\ptr{S}{U (\varrho\subtiny{0}{0}{S} \otimes \tau\subtiny{0}{0}{M}(\beta,H\subtiny{0}{0}{M})) U^\dagger} - \tau\subtiny{0}{0}{M}(\beta,H\subtiny{0}{0}{M})\}} \notag \\ 
    &\text{subject to:} \quad\quad &&\varrho\subtiny{0}{0}{S}^\prime = \ptr{M}{U (\varrho\subtiny{0}{0}{S} \otimes \tau\subtiny{0}{0}{M}(\beta,H\subtiny{0}{0}{M})) U^\dagger}.
\end{align}
The free variables here are $U$ and $H\subtiny{0}{0}{M}$, thereby constituting a double optimisation problem. Although the cost function to be minimised is linear in $U$, it is manifestly nonlinear in $H\subtiny{0}{0}{M}$, which appears once as a linear factor and once implicitly as an exponential term via $\tau\subtiny{0}{0}{M}(\beta,H\subtiny{0}{0}{M})$. Such a highly nonlinear form means that many standard optimisation methods, such as those used for linear or quadratic problems, are not suitable.

\begin{figure}
    \centering
\includegraphics[width=0.98\linewidth]{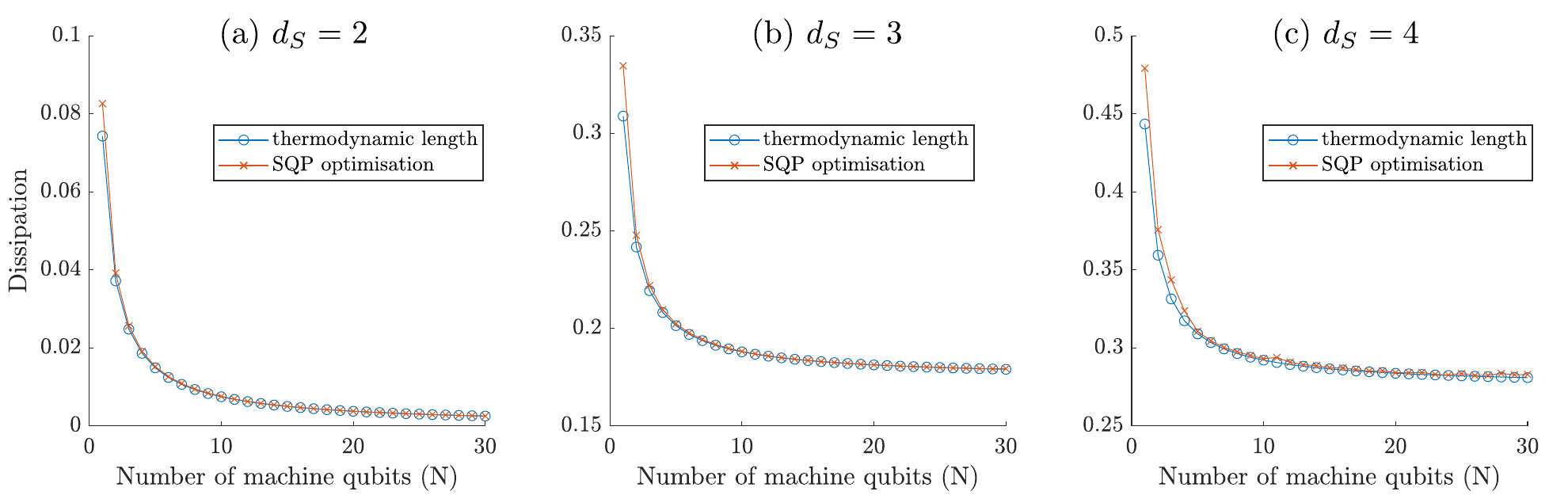}
    \caption{\textit{Comparison of the thermodynamic-length protocol with numerical optimisation}. We plot the dissipation $\beta \Delta E\subtiny{0}{0}{M} -\widetilde{\Delta} S\subtiny{0}{0}{S}$ as a function of the number of machine qubits $N$ for (a) qubit, (b) qutrit and (c) ququart target systems. While our numerical SQP approach is not guaranteed to yield the optimal solution, we readily see that the values obtained are very close to those of the thermodynamic-length protocol.}
    \label{fig:numerics}
\end{figure}

In order to proceed further, note that we can recast the optimisation problem~\eqref{eq::optimisationdensityoperators} into a simpler vectorised form as follows. Let $\vec{\omega}\subtiny{0}{0}{M} := (\omega_0, \hdots, \omega_{d\subtiny{0}{0}{M}-1})$ be the $1 \times d\subtiny{0}{0}{M}$ row vector of energy eigenvalues and $\vec{\nu}\subtiny{0}{0}{SM} := (\nu_0,\hdots,\nu_{d\subtiny{0}{0}{S} d\subtiny{0}{0}{M}-1}) = (\lambda_0 \mu_0, \lambda_0 \mu_1, \hdots , \lambda_0 \mu_{d\subtiny{0}{0}{M}-1}, \lambda_1 \mu_0, \hdots, \lambda_{d\subtiny{0}{0}{S}-1} \mu_{d\subtiny{0}{0}{M}-1}) =: \vec{\lambda}\subtiny{0}{0}{S} \otimes \vec{\mu}\subtiny{0}{0}{M}$ be the $d\subtiny{0}{0}{S} d\subtiny{0}{0}{M} \times 1$ column vector of initial global state spectrum, where $\vec{\lambda}\subtiny{0}{0}{S} := (\lambda_0,\hdots,\lambda_{d\subtiny{0}{0}{S}-1})$ and $\vec{\mu}\subtiny{0}{0}{M} := (\mu_0,\hdots,\mu_{d\subtiny{0}{0}{M}-1})$ describe the initial spectra of the system and machine states, respectively. We also define $T\subtiny{0}{0}{S}$ to be the $d\subtiny{0}{0}{M} \times d\subtiny{0}{0}{S} d\subtiny{0}{0}{M}$ (fixed) matrix that corresponds to tracing out the system degrees of freedom, and $T\subtiny{0}{0}{M}$ to be the $d\subtiny{0}{0}{S} \times d\subtiny{0}{0}{S} d\subtiny{0}{0}{M}$ (fixed) matrix that corresponds to tracing out the machine degrees of freedom. The matrix $A$, where $A\subtiny{0}{0}{ij}:=\vert U\subtiny{0}{0}{ij}\vert^2$, is a special form of doubly stochastic $d\subtiny{0}{0}{S} d\subtiny{0}{0}{M} \times d\subtiny{0}{0}{S} d\subtiny{0}{0}{M}$ matrix that corresponds to the action of a unitary operation that acts sequentially on the target system and each of the machine subsystems. With this, the optimisation problem~\eqref{eq::optimisationdensityoperators} takes the form:
\begin{align}\label{eq::optimisationvector}
    &\text{given} \quad\quad &&\vec{\lambda}\subtiny{0}{0}{S}, \,\vec{\lambda}\subtiny{0}{0}{S}^\prime, \, \beta \notag \\
    &\text{minimise} \quad\quad && \vec{\omega}\subtiny{0}{0}{M} \cdot (T\subtiny{0}{0}{S} \cdot A \cdot \vec{\nu}\subtiny{0}{0}{SM} - \vec{\nu}\subtiny{0}{0}{SM}) \notag \\ 
    &\text{subject to} \quad\quad &&\vec{\lambda}\subtiny{0}{0}{S}^\prime = T\subtiny{0}{0}{M} \cdot A \cdot \vec{\nu}\subtiny{0}{0}{SM}.
\end{align}
This non-linear optimisation problem can be heuristically solved using a \emph{sequential quadratic programming} (\textbf{SQP}) algorithm. Although this approach is not guaranteed to yield the global optimum, in many cases it provides a very good solution, as we highlight in Fig.~\ref{fig:numerics}. Notably, our numerical analysis suggests that the thermodynamic-length protocol may indeed be optimal for general $d$-dimensional systems: We observe that as $N$ increases and thus first-order corrections in Eq.~\eqref{eq::coherentthermodynamiclength} become dominant, the dissipation obtained in the protocol found by numerical optimisation converges to that found using the thermodynamic-length approach.

In the special case of a qubit target interacting with qubit machines, we can nonetheless analytically prove optimality of the sequence of swap operations, as we now demonstrate.\\[-2mm]

\textbf{Qubit-qubit case.---}One of the key difficulties in analytically solving this optimisation problem lies in characterising the conditions for which a given transformation $\varrho\subtiny{0}{0}{S} \mapsto \varrho\subtiny{0}{0}{S}^\prime$ is possible. Fortunately, for the case of a qubit target interacting with qubit machines $d\subtiny{0}{0}{S} = d\subtiny{0}{0}{M} = 2$, we can precisely characterise such states via the following lemma. 
\begin{lem}[Refs.~\cite{Clivaz_2019L,Clivaz_2019E}]
    For given qubit states $\varrho\subtiny{0}{0}{S}$ and $\sigma\subtiny{0}{0}{M}$ such that $\varrho\subtiny{0}{0}{S} \prec \sigma\subtiny{0}{0}{M}$, all achievable marginals from a global unitary evolution $U$ satisfy 
\begin{align}
    \frac{\mathbbm{1}}{2} \,\prec \, \mathrm{tr}\subtiny{0}{0}{S/M}[U(\varrho\subtiny{0}{0}{S} \otimes \sigma\subtiny{0}{0}{M}) U^\dagger] \,\prec \, \sigma\subtiny{0}{0}{M}.
    \label{eq:maj qubit}
\end{align}
\end{lem}
\noindent Since any pair of qubit states admits a majorisation relation between them, the above condition fully characterises the set of achievable transformations. Thus, a qubit-to-qubit transformation $\varrho\subtiny{0}{0}{S} = \diag(\lambda,1-\lambda) \mapsto \varrho\subtiny{0}{0}{S}^\prime = \diag(\lambda^\prime,1-\lambda^\prime)$, where $\lambda^\prime\geq \lambda$, is possible via Eq.~\eqref{eq::optimality-1} if and only if 
\begin{align}\label{eq::qubitmajorisation}
    \varrho\subtiny{0}{0}{S}^\prime \prec \tau\subtiny{0}{0}{M}(\beta,H\subtiny{0}{0}{M}).
\end{align}
There is an entire family of thermal states that majorise $\varrho\subtiny{0}{0}{S}$, namely any that have ground-state population larger than that of the final state; from any of these, one can achieve the desired $\varrho\subtiny{0}{0}{S}^\prime$. Given that the target system and machine begin at the same temperature and both of their ground-state energies are set to zero without loss of generality, said family of thermal states is characterised by $\omega\subtiny{0}{0}{M} \geq \omega\subtiny{0}{0}{S}$, where $\omega\subtiny{0}{0}{X}$ denotes the energy gap of the system with Hamiltonian $H\subtiny{0}{0}{X}$.

The question is now reduced to: \emph{What is the optimal global operation (in the sense of minimising the dissipated heat) to apply to a thermal machine state of this form in order to achieve the desired transformation?} Since creating correlations and coherences both incur energy costs~\cite{Huber_2015,BruschiPerarnauLlobetFriisHovhannisyanHuber2015,Bakhshinezhad_2019}, we must consider the parameterised family of probabilistic swap operations, i.e.,
\begin{align}
    U = q\,\mathbb{S} + (1-q) \openone,
\end{align}
where $q \in [0,1]$ and $\mathbb{S}$ and $\openone$ denote the swap and identity operators, respectively. This is the most general form of operations between diagonal states that do not create any correlations or coherences in the marginals. In terms of the vectorised formulation of the problem [see~\eqref{eq::optimisationvector}], this family of operations corresponds to the matrix
\begin{align}
    A = \left[\begin{array}{cccc}
        1 & 0 & 0 & 0 \\
        0 & 1-q & q & 0 \\
        0 & q & 1-q & 0 \\
        0 & 0 & 0 & 1
    \end{array}\right].
\end{align}
Let the initial state of the target system correspond to $\vec{\lambda}\subtiny{0}{0}{S} = (\lambda,1-\lambda)$ and that of the machine to $\vec{\mu}\subtiny{0}{0}{M} = (\mu,1-\mu)$, where $\mu = (1+e^{-\beta \omega\subtiny{0}{0}{M}})^{-1}$. The general form of the global vector $\vec{\nu}\subtiny{0}{0}{SM}$ after the probabilistic swap operation is
\begin{align}
    \vec{\nu}^{\,\prime}\subtiny{0}{0}{SM} = \left[\begin{array}{c}
        \lambda \mu \\
        (1-q) \lambda (1-\mu) + q (1-\lambda) \mu \\
        q \lambda (1-\mu) + (1-q)(1-\lambda) \mu \\
        (1-\lambda) (1-\mu)
    \end{array}\right].
\end{align}
The final ground-state population of the target $\lambda^\prime = \vec{\nu}^{\,\prime}_0 + \vec{\nu}^{\,\prime}_1$ is thus 
\begin{align}
    \lambda^\prime = q \mu + (1-q) \lambda.
\end{align}
The dissipated heat is given by the machine energy gap multiplied by the change in the excited-state population of the machine
\begin{equation}
\Delta E\subtiny{0}{0}{M} = \omega\subtiny{0}{0}{M} (\mu -\mu^\prime) = \omega\subtiny{0}{0}{M}[\mu - \lambda \mu - q \lambda (1-\mu) - (1-q)(1-\lambda)\mu] = \omega\subtiny{0}{0}{M}[q (\mu - \lambda)] =: \omega\subtiny{0}{0}{M} \Delta p,
\end{equation}
where $\Delta p$ represents the population exchanged via the transformation. Note that this quantity, which is a function of both the probability of swapping $q$ and the machine energy gap (implicitly via $\mu$), is fixed by the given initial and final states of the target system. Thus, minimising the energy cost per population exchange for a given transformation $\varrho\subtiny{0}{0}{S} \mapsto \varrho\subtiny{0}{0}{S}^\prime$, amounts to minimising $\omega\subtiny{0}{0}{M}$. In other words, the transformation should be implemented using the smallest energy gap in the machine; based on Eq.~\eqref{eq:maj qubit}, the smallest gap that permits the desired transformation is precisely $ \tau\subtiny{0}{0}{M}(\beta,\omega\subtiny{0}{0}{M})=\tau\subtiny{0}{0}{S}(\beta_{\mathrm{f}}, \omega\subtiny{0}{0}{S})$. This gives us the optimal initial state of the machine. It is straightforward to see that the desired transformation of the target system is then obtained via the unitary that performs a complete swap, i.e., $q=1$.


\subsection{Coherent-control scenario (explicit geodesic solution for the case of a qubit target system)}
\label{app::coherentqubitoptimal}

We now present the explicit geodesic solution for the case where the system is a qubit (i.e., $d\subtiny{0}{0}{S} = 2$). In this case we can parameterise the Hamiltonian from Eq.~\eqref{eq:ham_param} as $H_t = H_0 + \xi_t X$, where $\xi_t \in [0,1]$ is a continuous function of $t$ and $X$ is a Hermitian operator. Without loss of generality we can assume $H_0$ and $X$ to be of the form
\begin{equation}
\label{eq:qubit_example_param}
   H_0 = 0 \hspace{10pt} \textup{and}\hspace{10pt}  X = \begin{bmatrix}
0 & 0 \\
0 & 1
\end{bmatrix},
\end{equation}
which greatly simplifies the analysis. 
Our goal is to find a family of functions $\xi_t$ that minimises the thermodynamic-length functional $\mathcal{L}$ from Eq.~\eqref{eq:thermo_length_def}. More specifically, using the parameterisation from Eq.~\eqref{eq:qubit_example_param}, the thermodynamic length $\mathcal{L}$ can be written as
\begin{align}
    \mathcal{L} &= \beta^2 \int_{0}^1 \sqrt{\dot{\xi_t}  m(\xi_t) \dot{\xi_t}} \,\text{d}t,
\end{align}
where $m(\xi_t)$ is a metric given by $m(\xi_t) = \tr{X^2\tau_t(\xi_t)} - \tr{X \tau_t(\xi_t)}^2 = e^{\beta \xi_t}{(1+e^{\beta \xi_t})^{-2}}$, where 
\begin{align}
    \tau_t(\xi_t) := e^{-\beta (H_0+\xi_t X)}/\mathrm{tr}[e^{-\beta (H_0+\xi_t X)}].
\end{align}
The optimal path can now be found by solving the Euler-Lagrange equation 
\begin{equation}
    \label{eq:euler_lagrange_qubit}
    \frac{\partial \mathcal{L}}{\partial \xi_t} = \frac{\text{d}}{\text{d} t} \frac{\partial \mathcal{L}}{\dot{\xi}_t}.
\end{equation}
This leads to an equation of the form $\ddot{\xi_t} + \Gamma_t \dot{\xi}^2_t = 0$, where $\Gamma_t$ is a Christoffel symbol. In our case, we can explicitly write 
\begin{equation}
    \Gamma_t = \frac{1}{2 m(\xi_t)} \frac{\partial m(\xi_t)}{\partial \xi_t}. 
\end{equation}
Using our expression for the metric we can solve the above equation for $\Gamma_t$, which leads to $\Gamma_t = -\tfrac{1}{2}\tanh(\tfrac{\xi_t}{2})$. With this, we can now solve Eq.~\eqref{eq:euler_lagrange_qubit} for $\xi_t$, which yields
\begin{align}
    \xi_t = 2 \, \mathrm{arcsinh}\left[\tan\left(\frac{c_1(t + c_2)}{2}\right)\right],
    \label{eq; optimal gap structure qubit}
\end{align}
where $c_1, c_2 \in \mathbb{R}$ are constants that depend on the boundary conditions of the cooling scheme, i.e., the initial and final temperature of the target system. For concreteness, suppose we choose $\beta$ and $\beta_{\mathrm{f}}$ such that the initial state of the system is given by $\varrho\subtiny{0}{0}{S} = \text{diag}(1-p, p)$ and the final one by $\varrho'\subtiny{0}{0}{S} = \text{diag}(1-q, q)$, where $p, q \in [0, 1]$\@. Due to our discretisation, the time parameter is given by $t = i/N$, where $i$ labels the machine qubit $M_i$ and $N$ is the total number of machine qubits. In this case, the boundary conditions for the trajectory $\xi_t$ are $\xi_0 = - \frac{1}{\beta} \log \left[\frac{p}{1-p}\right]$ and $\xi_1 = - \frac{1}{\beta} \log \left[\frac{q}{1-q}\right]$\@.

\subsection{Coherently cooling a qubit target system with a high-dimensional machine}
\label{app::highdimensionalmachine}

We have demonstrated the effectiveness of employing the thermodynamic length to derive minimally-dissipative processes in cases where the system is swapped with a sequence of machines of the same dimension. We now move to investigate the role of the machine's dimension and the system-machine interactions during the cooling protocol. We analyse a qubit target system and retain the assumption of a Markovian collision model regarding the cooling scheme, but now consider more complex machines: first described by a sequence of independent qu\textit{d}it machines before extending to permit an \textit{interacting} $N$-qubit machine, where $d\subtiny{0}{0}{M} = 2\suptiny{0}{0}{N}$. This example highlights that in the case of cooling with complex machines -- be they high-dimensional systems or interacting qubits -- that a sequence of swap operations no longer provides the optimal cooling strategy. 

Consider a single step of a qubit target system $S$ interacting with a fresh qudit machine $M$ of dimension $d\subtiny{0}{0}{M} =: d$ and whose Hamiltonian is equally spaced (with arbitrary frequencies $\omega\subtiny{0}{0}{M} $), i.e., $H\subtiny{0}{0}{M} = \sum_{n=0}^{d-1} n \, \omega\subtiny{0}{0}{M} \ketbra{n}{n}$. We assume that the machine begins in a thermal state with inverse temperature $\beta$, i.e., $\tau\subtiny{0}{0}{M}(\beta)= \sum_{n=0}^{d-1} p_{n} \ketbra{n}{n}$, where $p_{n} = \frac{e^{-\beta n \omega\subtiny{0}{0}{M}} (1-e^{-\beta \omega\subtiny{0}{0}{M}})}{1-e^{-\beta d \omega\subtiny{0}{0}{M}}}$. Since we consider a qubit target system, its initial state can always be described as a thermal state in terms of its energy gap as long as it is diagonal in the energy basis, i.e., $\tau\subtiny{0}{0}{S}(\beta\subtiny{0}{0}{S})= \sum_{m=0}^{1} q_{m} \ketbra{m}{m}$, where $q_{m} = \frac{e^{-m\beta\subtiny{0}{0}{S} \omega\subtiny{0}{0}{S}}}{1+e^{-\beta\subtiny{0}{0}{S} \omega\subtiny{0}{0}{S}}}$. Finally, we consider applying the following form of interaction unitary
\begin{align}
U\subtiny{0}{0}{{SM}} = \sum_{n=0}^{d-2} (\ketbra{0, n+1}{1,n} + \ketbra{1,n}{0, n+1}) + \ketbra{0, 0}{0,0} + \ketbra{1,d-1}{1, d-1}.
\end{align}
After applying such a unitary the resulting global state is given by
\begin{align}\label{app::eq-highdimtransformation}
\varrho^\prime\subtiny{0}{0}{{SM}} &= U\subtiny{0}{0}{{SM}} [\tau\subtiny{0}{0}{S}(\beta\subtiny{0}{0}{S}) \otimes \tau\subtiny{0}{0}{M}(\beta)] U^{\dagger}\subtiny{0}{0}{{SM}} \nonumber\\
&= \sum_{n=0}^{d-2} \bigg(q_1 p_{n} \ketbra{0,n+1}{0,n+1} + q_0 p_{n+1} \ketbra{1,n}{1,n} \bigg) + q_0 p_{0} \ketbra{0,0}{0,0} + q_1 p_{d-1} \ketbra{1,d-1}{1,d-1}.
\end{align}
Note that no coherence is generated by this transformation and that both the machine and target system remain diagonal in their local energy eigenbases. Since the interaction only involves the qubit target and virtual qubits of the machine with the same energy gap $\omega\subtiny{0}{0}{M}$, we can obtain the heat dissipated by the machine $\Delta E\subtiny{0}{0}{M} = \omega\subtiny{0}{0}{M} \Delta q_0$ in terms of the change in the ground-state population of the target system $\Delta q_0 := q_0^{\prime} - q_0$. To obtain $q_0^{\prime}$, we calculate
\begin{align}
q_0^{\prime} &= q_0 p_0 + \sum_{n=0}^{d-2} q_1 p_{n} = \frac{(1-e^{-\beta \omega\subtiny{0}{0}{M}})}{(1-e^{-\beta d \omega\subtiny{0}{0}{M}})(1+e^{-\beta\subtiny{0}{0}{S} \omega\subtiny{0}{0}{S}})}  \left(1+\frac{e^{-\beta\subtiny{0}{0}{S} \omega\subtiny{0}{0}{S}}(1-e^{-\beta (d-1) \omega\subtiny{0}{0}{M}})}{1-e^{-\beta \omega\subtiny{0}{0}{M}}}\right).
\end{align}
Thus, $\Delta q_0$ is given by
\begin{equation}
\Delta q_0 = \frac{(e^{\beta (d-1) \omega\subtiny{0}{0}{M}} - 1) (e^{\beta \omega\subtiny{0}{0}{M}} - e^{\beta\subtiny{0}{0}{S}\omega\subtiny{0}{0}{S}})}{(e^{\beta d \omega\subtiny{0}{0}{M}} - 1) (1 + e^{\beta\subtiny{0}{0}{S} \omega\subtiny{0}{0}{S}})}.
\end{equation}
This equation indicates that if $\beta \omega\subtiny{0}{0}{M} \geq \beta\subtiny{0}{0}{S} \omega\subtiny{0}{0}{S}$, the protocol will cool the target system; otherwise, the target will heat up. In addition, the above expression allows us to analyse the role of the machine dimension $d$, the increase of which can improve cooling (and heating) performance, as we now demonstrate. 

Consider the situation where $\beta\subtiny{0}{0}{S} = \beta$ and $\omega\subtiny{0}{0}{M} \geq \omega\subtiny{0}{0}{S}$. In the limit of $d \to \infty$ (i.e., cooling a qubit target with a harmonic oscillator machine), the ratio between the ground $q_0^{\prime}$ and excited $q_1^{\prime}$ state populations of the target system after the transformation described by Eq.~\eqref{app::eq-highdimtransformation} is given by
\begin{equation}
\frac{q_0^{\prime}}{q_1^{\prime}} = e^{\beta \omega\subtiny{0}{0}{M}} + e^{\beta (\omega\subtiny{0}{0}{M} - \omega\subtiny{0}{0}{S})} - 1 =: e^{\beta \nu},
\label{eq: final population ratio}
\end{equation}
where $\nu \geq \omega\subtiny{0}{0}{M}$. Thus, although all of the virtual qubits of the machine that interact with the system correspond to the energy gap $\omega\subtiny{0}{0}{M}$, the target system can reach a state colder than that set by this virtual temperature, namely that corresponding to the parameter $\nu$. Since the target system is cooled, the heat dissipated by the machine is positive, with the energy cost per population change determined by 
\begin{equation}
\frac{\Delta E^{\text{h.o.}}\subtiny{0}{0}{M}}{\Delta q_0} = \omega\subtiny{0}{0}{M}.
\label{eq:per unit}
\end{equation}
To summarise, via the transformation in Eq.~\eqref{app::eq-highdimtransformation}, the harmonic oscillator machine allows the target system to be cooled beyond the temperature of the virtual machine qubits [see Eq.~\eqref{eq: final population ratio}] but at an energy cost per population exchange that corresponds to the virtual machine qubits [see Eq.~\eqref{eq:per unit}]. 

Now, if one wished to cool the target system to the same extent using a \textit{qubit} machine, then they would require a machine qubit with at least energy gap $\nu_i$. In this case, the energy cost of cooling would be larger than that derived above using the harmonic oscillator machine, since
\begin{equation}
\frac{\Delta E^{\text{qubit}}\subtiny{0}{0}{M}}{\Delta q_0} = \nu \geq \omega\subtiny{0}{0}{M} = \frac{\Delta E^{\text{h.o.}}\subtiny{0}{0}{M}}{\Delta q_0}.
\label{eq:cost comparison}
\end{equation}
\noindent This shows that heat dissipation can be minimised by using a higher-dimensional machine.

Finally, by considering $d= \, 2\suptiny{0}{0}{N}$, it is clear that there is an advantage from using a machine comprising interacting qubits.  


\section{Cooling Using Energy Gaps from the Gaussian Orthogonal Ensemble}

Here, we investigate the energy cost of cooling with a Markovian collision model setting where the (local) energy gap structure of an $N$-qubit machine is characterised by an ordered \textit{Gaussian Orthogonal Ensemble} (\textbf{GOE}) \cite{Livan_2018}. The GOE---well-known in the statistical mechanics community---provides a framework for modelling energy levels in complex quantum systems. Due to its distinctive energy level repulsion properties, this ensemble is particularly relevant for studying systems that thermalise \cite{weidenmuller2024}.

In what follows, we describe two different cooling protocols inspired by the GOE. First, we consider a protocol that we refer to as the \emph{Eigenvalue GOE} protocol, in which the machine gaps are directly sampled from the GOE distribution before being appropriately sorted and rescaled. Second, we consider a protocol that we call \emph{Spacing GOE}, in which machine gaps are sampled from the nearest-neighbour GOE distribution. In this case, we first sample eigenvalues from the GOE distribution, sort them non-increasingly, and then calculate the nearest-neighbour energy gap spacings. In what follows we describe sampling procedure for both protocols in detail. \\

\textbf{Sampling matrices from the GOE.---}Let us begin by denoting the gaps of $N$ machine qubits with Hamiltonians $\{H_{M_i}\}_i$ as $\{\xi_{i}\}_{i=1}^N$. Both protocols that we consider here rely upon first generating matrices from the Gaussain orthogonal ensemble (GOE) and then computing eigenvalues. 

\begin{enumerate}
    \item Generate $m$ random matrices $\{M_i\}_{i=1}^m$ of dimension $N+1$ from the GOE. 
    \begin{enumerate}
        \item Generate $m \times (N+1)^2$ real numbers distributed randomly according to a standard normal distribution [i.e., the Gaussian distribution $\mathcal{N}(\mu, \sigma)$ with zero mean ($\mu = 0$) and unit variance ($\sigma = 1$)]. 
        \item Arrange the numbers in $m$ matrices $R_i$.
        \item Symmetrise the matrices to obtain 
        \begin{align}
            M_i := \frac{1}{\sqrt{2}} (R_i^{\textup{T}} + R_i).
        \end{align}
    \end{enumerate}
    \item  For each random matrix $M_i$, compute the sorted, normalised eigenvalues $\{\hat{\lambda}_k^{(i)}\}_{k=1}^{N+1}$. 
    \begin{enumerate}
        \item Compute the eigenvalues $\{\lambda_k^{(i)}\}_{k=1}^{N+1}$ of each matrix $M_i$ and normalise them via
        \begin{align}
            \tilde{\lambda}_{k}^{(i)} := \frac{1}{\sqrt{N+1}} \lambda_{k}^{(i)}.
        \end{align}
        \item Sort the normalised eigenvalues $\tilde{\lambda}_{k}^{(i)}$ in non-increasing order. Denote these sorted eigenvalues by $\hat{\lambda}_{k}^{(i)}$ such that $\hat{\lambda}_{k}^{(i)} \geq \hat{\lambda}_{k+1}^{(i)}$ for all $k \in \{1, \ldots, N+1\}$ and $i \in \{1, \ldots, m\}$. 
    \end{enumerate}

\end{enumerate}

\noindent At this point, the two procedures diverge. We first explain the Eigenvalue GOE protocol.\\

\noindent \textbf{Sampling machine energy gaps from GOE distribution (Eigenvalue GOE protocol).---}Continuing from above, we then:
\begin{enumerate}[3.]
    \item Compute the sorted average eigenvalues $\{\hat{\lambda}_{k}\}_{k=1}^N$ to construct the machine energy gaps $\{\xi_{k}\}_{k=1}^N$.
    \begin{enumerate}
        \item Compute the averages $\{\hat{\lambda}_{k}\}_{k=1}^N$ over all realisations, namely
        \begin{align}
            \hat{\lambda}_{k} := \frac{1}{m}\sum_{i=1}^m \hat{\lambda}_{k}^{(i)}.
        \end{align}
        \item Sort the average eigenvalues $\{\hat{\lambda}_{k}\}_{k=1}^N$ in non-increasing order.
        \item Rescale these average eigenvalues to reflect the boundary conditions of the cooling process, namely 
    \begin{align}
        \label{eq:goe_eig}
        \xi_{k} := x_0 \left(\frac{\hat{\lambda}_k - \hat{\lambda}_N}{\hat{\lambda}_1  - \hat{\lambda}_N}\right) - x_1 \left(\frac{\hat{\lambda}_k - \hat{\lambda}_1}{\hat{s}_1 - \hat{\lambda}_N}\right), \,\, \text{where} \,\, x_0 :=  -\frac{1}{\beta} \log \left(\frac{p}{1-p}\right) \,\, \text{and} \,\, x_1 := -\frac{1}{\beta} \log \left(\frac{q}{1-q}\right). 
    \end{align}
    In this way, the machine qubit gaps satisfy $\xi_1 = x_0$ and $\xi_N = x_1$, and hence the sequence of swaps transforms the system qubit from $\varrho\subtiny{0}{0}{S} = \text{diag}(1-p, p)$ to $\varrho'\subtiny{0}{0}{S} = \text{diag}(1-q, q)$. 
    \end{enumerate}

\end{enumerate}

\noindent We now return to the sorted eigenvalues of the GOE matrices (step 2), from which we present the Spacing GOE protocol. \\

\noindent \textbf{Sampling machine energy gaps from nearest-neighbour GOE distribution (Spacing GOE protocol).---}Continuing, we then:
\begin{enumerate}[$3^\prime$.]
    \item Calculate the average nearest-neighbour spacings $\{\tilde{s}_{k}\}_{k=1}^N$ to construct the machine energy gaps $\{\xi_{k}\}_{k=1}^N$.
    \begin{enumerate}
        \item Compute the nearest-neighbour spacings via the difference between adjacent eigenvalues in the set $\{ \hat{\lambda}_{k}^{(i)} \}$, namely
        \begin{align}
            s_{k}^{(i)} := \hat{\lambda}_{k+1}^{(i)} - \hat{\lambda}_{k}^{(i)} \qquad \text{for all}\quad i \in \{1, \ldots, m\}.
        \end{align}
        \item Normalise the nearest-neighbour spacings as
        \begin{align}
            \tilde{s}_{k}^{(i)} := \frac{1}{A_i} s_{k}^{(i)} \qquad \text{where} \quad A_i := \sum_{k=1}^N s_{k}^{(i)} 
        \end{align}
        \item Compute the averages $\{\tilde{s}_{k}\}_{k=1}^N$ of the nearest-neighbour spacings over all realisations, i.e.,
        \begin{align}
            \tilde{s}_{k} := \frac{1}{m}\sum_{i=1}^m \tilde{s}_{k}^{(i)}.
        \end{align}
        \item Sort the average spacings $\{\tilde{s}_{k}\}_{k=1}^N$ in non-increasing order. Denote these sorted spacings by $\hat{s}_k$ such that $\hat{s}_{k+1} \geq \hat{s}_{k}$.
        \item Rescale the spacings to reflect the boundary conditions of the cooling process, namely  
    \begin{align}
        \label{eq:goe_spacings}
        \xi_{k} := x_0 \left(\frac{\hat{s}_k - \hat{s}_N}{\hat{s}_1  - \hat{s}_N}\right) - x_1 \left(\frac{\hat{s}_k - \hat{s}_1}{\hat{s}_1 - \hat{s}_N}\right), \quad \text{where} \,\, x_0 :=  -\frac{1}{\beta} \log \left(\frac{p}{1-p}\right) \,\, \text{and} \,\, x_1 := -\frac{1}{\beta} \log \left(\frac{q}{1-q}\right). 
    \end{align}
    In this way, the machine qubit gaps satisfy $\xi_1 = x_0$ and $\xi_N = x_1$, and hence the sequence of swaps transforms the system qubit from $\varrho\subtiny{0}{0}{S} = \text{diag}(1-p, p)$ to $\varrho'\subtiny{0}{0}{S} = \text{diag}(1-q, q)$. 
    
    \end{enumerate}

\end{enumerate}

\begin{figure}[t]
    \centering
    \begin{minipage}{0.45\textwidth}
        \centering
        \includegraphics[width=\textwidth]{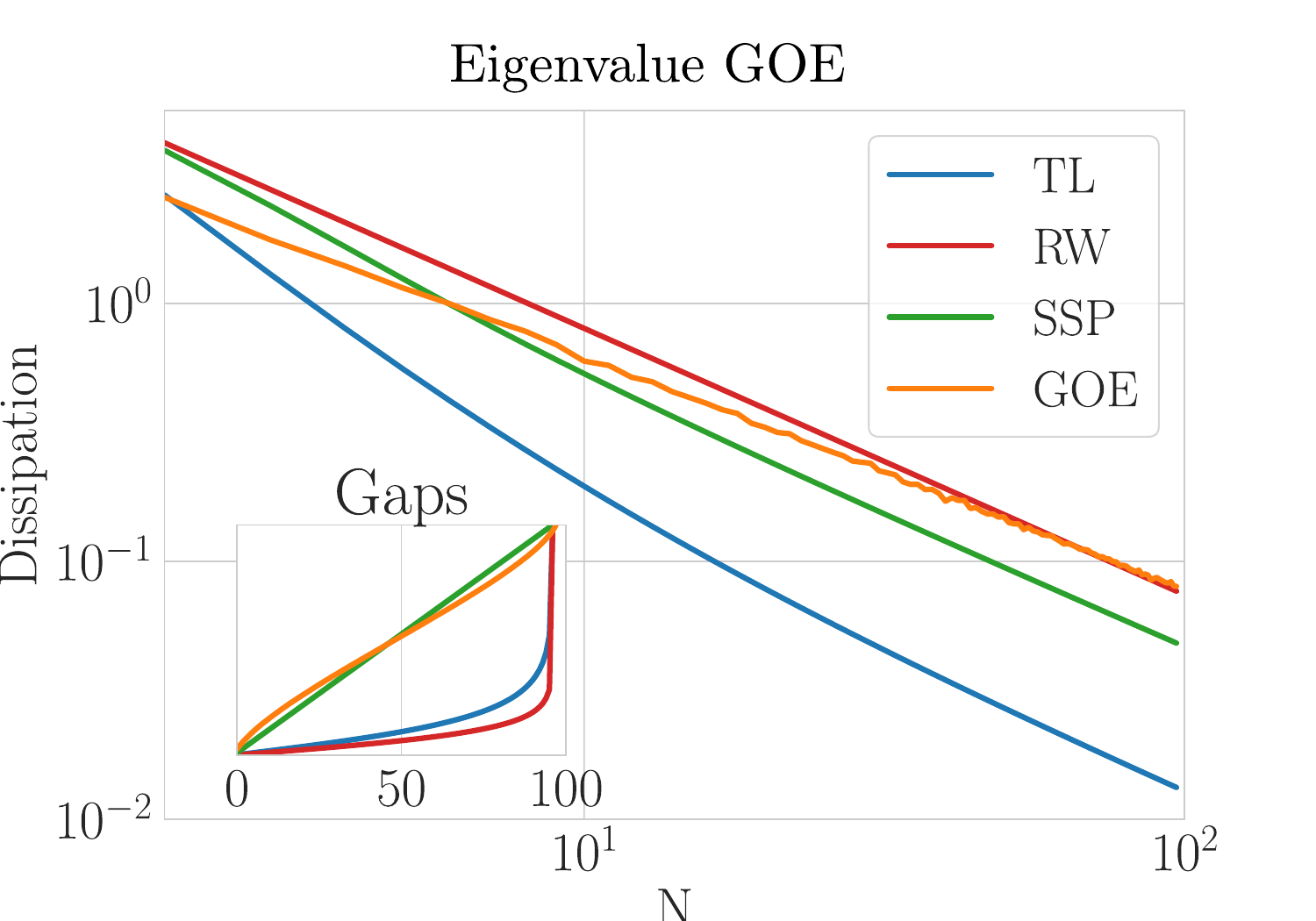} 
    \end{minipage}\hfill
    \begin{minipage}{0.45\textwidth}
        \centering
        \includegraphics[width=\textwidth]{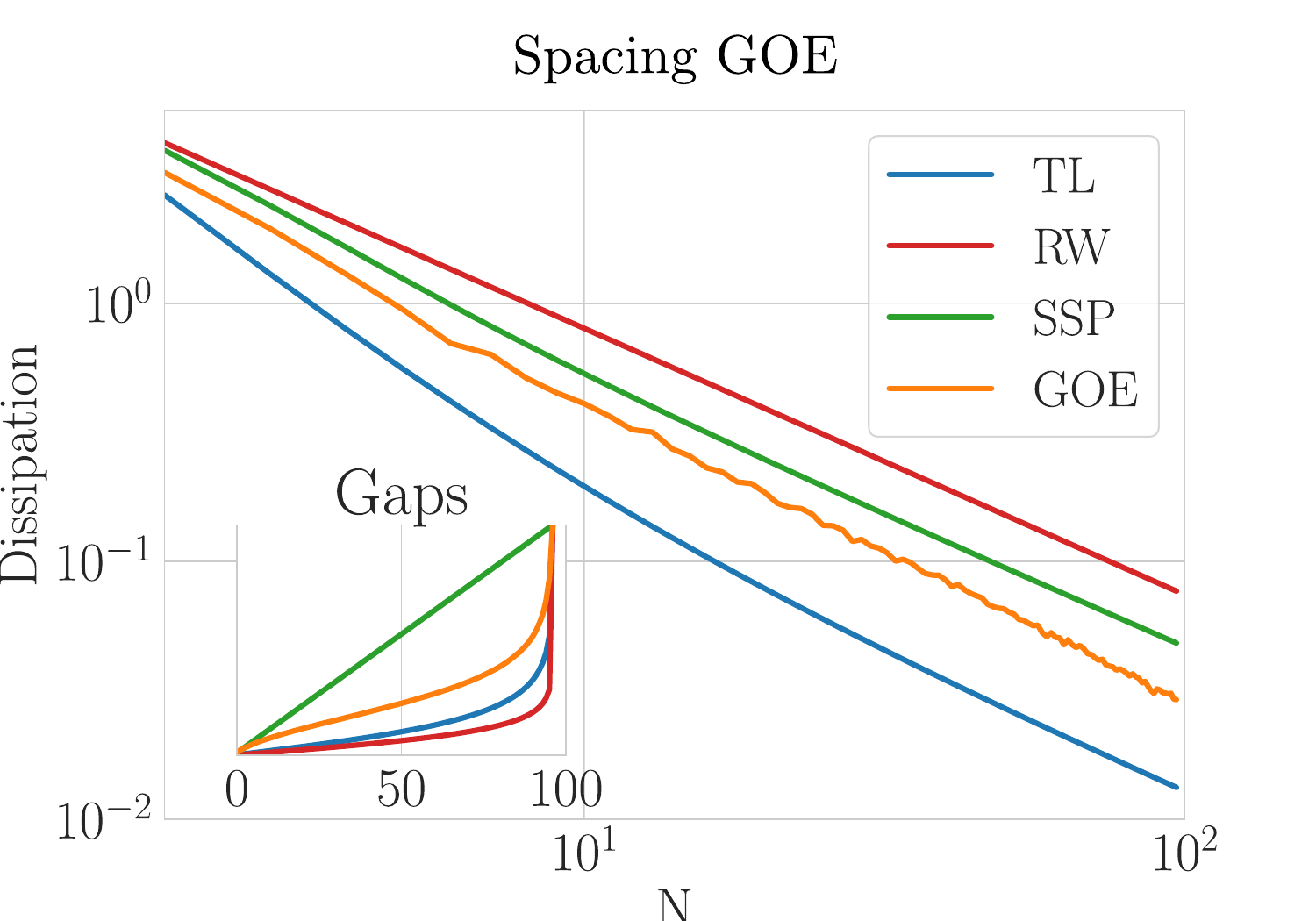} 
    \end{minipage}
    \caption{\textit{Cooling with Coherent Control.} Heat dissipation when coherently cooling a qubit target with energy gap $E\subtiny{0}{0}{S} = 1$ from $\beta = 10^{-8}$ to $\beta_{\mathrm{f}} = 10$ (i.e., $\lambda = 10^9$) via sequential bipartite interactions with $N$ machine qubits. We compare four procedures that depend on the energy-level structure $\{H_{M_i}\}_i$, whose gaps are depicted in the inset for $N = 100$. TL is the optimal protocol deduced from the thermodynamic length (see Theorem~\ref{thm::thermodynamiclengthcoherent}); RW corresponds to the protocol from Ref.~\cite{Reeb_2014} where machine energy gaps change linearly; SSP corresponds to the protocol from Ref.~\cite{skrzypczyk2014work} where the machine qubits' excited state populations change linearly. The left panel depicts the Eigenvalue GOE protocol [see Eq. \eqref{eq:goe_eig}]; the right panel depicts the Spacing GOE protocol [see Eq. \eqref{eq:goe_spacings}]. \label{fig_app_goe} }
\end{figure}

\noindent In Fig.~\ref{fig_app_goe}, we compare the energy cost of cooling (as quantified by dissipation) for the Eigenvalue GOE (left plot) and Spacing GOE (right plot) procedures with other cooling protocols discussed in the main text. Interestingly, we observe that the Spacing GOE protocol is quite efficient, even outperforming the RW and SSP cooling procedures previously discussed throughout the literature. As expected, the cooling protocol based on the thermodynamic length (TL) outperforms all other procedures.   

\section{Incoherent-control scenario (Proof of Theorem~\ref{thm::thermodynamiclengthincoherent})}\label{app::thermodynamiclengthincoherent}

We now move to analyse the previously discussed question with respect to the incoherent-control setting. The efficient incoherent cooling scheme that we present is formalised in Theorem~\ref{thm::thermodynamiclengthincoherent} of the main text, which provides a bound for the heat drawn from the hot bath. Additionally, we can make a similar statement for the heat dissipated into the cold bath. The restated theorem below encompasses both results.

\begin{thm}\label{thm::thermodynamiclengthIncoherentqudit}
In the incoherent-control setting, the minimum energy cost for cooling a qudit system with Hamiltonian $H$ from a thermal state $\tau\subtiny{0}{0}{S}(\beta, \, H)$ to  $\tau\subtiny{0}{0}{S}( \lambda  \beta, H)$, by using particular tripartite (energy-conserving) interactions between the system and two fresh qudit machines at inverse temperature $\beta$ (cold) and $\beta\subtiny{0}{0}{H} \leq \beta$ (hot), respectively, and with arbitrary Hamiltonians with $N$$<\infty$ distinct energy gaps but in the limit of infinite steps, is given by
\begin{align}
    \beta\Delta E\subtiny{0}{0}{C} =\, -(\gamma+1) \,\Delta S\subtiny{0}{0}{S} + \,\frac{ (\gamma+1) }{2 N} (\mathcal{L}^*)^2 +\gamma \beta\, \Delta E\subtiny{0}{0}{S}+ \mathcal{O}(N^{-2}),
    \label{eq: 3q incoh energy cost C}
\end{align}
where $\gamma:=\frac{\beta\subtiny{0}{0}{H}}{\beta-\beta\subtiny{0}{0}{H}}$, and $\mathcal{L}^*$ is the thermodynamic length computed for the optimal trajectory (geodesics) $\{\xi_i^*(t)\}_{i=1}^N$ which is the solution to Eq.~\eqref{eq:app-euler-lagrange}. Since the protocol considered is energy conserving, one can write $\Delta E\subtiny{0}{0}{C}$ in Eq.~\eqref{eq: 3q incoh energy cost C} in terms of the energy changes of $S$ and $H$, i.e., $\Delta E\subtiny{0}{0}{C}=\,-\Delta E\subtiny{0}{0}{S}-\, \Delta E\subtiny{0}{0}{H} $, from which it follows that
\begin{equation}\label{eq::appthmincoherenthot}
    \Delta F\subtiny{0}{0}{S}^\beta\,+\eta \,\Delta E\subtiny{0}{0}{H} = \,-\frac{1}{2N\beta} (\mathcal{L}^{*})^2 +\mathcal{O}(N^{-2}),
\end{equation}
where $\Delta F\subtiny{0}{0}{X}^\beta := \Delta E\subtiny{0}{0}{X}- \beta^{-1} \Delta S\subtiny{0}{0}{X}$ and $\eta := 1 - \beta\subtiny{0}{0}{H}/\beta= (\gamma+1)^{-1}$\@. For qubit target systems interacting with qubit hot and cold machines, the interactions considered are optimal in general.
\end{thm}

\noindent As a side remark, note that when $\gamma \to 0$, i.e., in the limit of an infinite-temperature hot bath, the result above [namely Eq.~\eqref{eq: 3q incoh energy cost C}] reduces to that of the coherent-control setting [i.e., Eq.~\eqref{eq::coherentthermodynamiclength}]. This reflects the intuition that an infinitely hot bath can be considered to be a source of infinite work. However, even in this limit, there is a crucial distinction between the two settings: it is also important to consider the rate at which the target state is cooled; as $\gamma\to 0$, the cooling speed in the incoherent protocol also goes to zero. 

\begin{proof}
We begin with a family of machine Hamiltonians and energy-conserving unitaries that can cool the target system. For the described protocol, we explicitly calculate the energy drawn from the hot bath (and that dissipated into the cold bath), which we then bound in terms of the thermodynamic length. We subsequently demonstrate optimality when all systems are qubits.

In general, any incoherent-control unitary $\widetilde{U}$ must satisfy the condition of energy conservation $[\widetilde{U}, H\subtiny{0}{0}{S} + H\subtiny{0}{0}{C} + H\subtiny{0}{0}{H}] = 0$\@. In order to satisfy this condition, the interaction Hamiltonian that generates this unitary must commute with the local Hamiltonians. In other words, $\widetilde{U}$ can be decomposed into unitaries $\widetilde{U} = \bigotimes_k U_{\text{deg}}\suptiny{0}{0}{(k)} \oplus \mathbbm{1}_{\text{rest}}$ that only act non-trivially in the degenerate subspaces of the total Hamiltonian $H := H\subtiny{0}{0}{S} + H\subtiny{0}{0}{C} + H\subtiny{0}{0}{H}$\@. Additionally, for the task of cooling the target, it is worth noting that the local Hamiltonians cannot be chosen arbitrarily. For instance, in the qubit case, choosing the energy gaps $E\subtiny{0}{0}{S}$ and $E\subtiny{0}{0}{C}$ constrains the energy gap of the hot machine to fulfill $E\subtiny{0}{0}{H}=\, \vert E\subtiny{0}{0}{C}-E\subtiny{0}{0}{S}\vert$. Otherwise, there is no non-trivial unitary that commutes with the sum of the local Hamiltonians, since there would be no degenerate subspace.

We then consider the following energy structures, where the Hamiltonians of the cold and hot baths are scaled versions of that of the target system,
\begin{equation}
 H\subtiny{0}{0}{C_n}=\left( \left( \gamma+1 \right) \lambda_n -\,\gamma \right) H\subtiny{0}{0}{S} \quad\quad \text{and} \quad\quad H_{H_n}= \left(\gamma+1\right) \left(\lambda_n -1\right) H\subtiny{0}{0}{S}. 
 \label{eq:qudit hamiltonian}
\end{equation}  
For such a structure of machine Hamiltonians, we will consider the specific class of energy-conserving unitaries that can be generated via interaction Hamiltonians of the form
\begin{equation}\label{eq::incoherentquditinteraction}
    \widetilde{H}_{\mathrm{int}} = \sum_{i=0}^{d-2} (\ketbra{i,\,i+1,\,i }{i+1,\,i,\,i+1}+\ketbra{i+1,\,i,\,i+1}{i,\,i+1,\,i }).
\end{equation}
The unitaries generated from such interactions act non-trivially on $d-1$ distinct two-dimensional orthogonal subspaces, where the $i^{\textup{th}}$ such subspace is spanned by the vectors $\{\ket{i,\,i+1,\, i }\subtiny{0}{0}{SCH},\, \ket{i+1,\, i,\,i+1}\subtiny{0}{0}{SCH}\}$\@. This operation represents an exchange between the subspace of the target system spanned by $\{\ket{i}\subtiny{0}{0}{S},\ket{i+1}\subtiny{0}{0}{S}\}$ and the virtual subspace of the machine spanned by $\{ \ket{i+1,\,i}\subtiny{0}{0}{CH},\ket{i,\,i+1}\subtiny{0}{0}{CH} \}$. Since such subspaces contain total populations strictly less than one, these exchanges with the virtual machine qubits must be repeated infinitely many times in principle, in order for the excited population $p^{(i+1,i)}\subtiny{0}{0}{CH}$ to be swapped with that of the lower-energy eigenstate of the system $p^{(i)}\subtiny{0}{0}{S}$~\cite{Clivaz_2019L,Clivaz_2019E}. Nonetheless, for a large number of repetitions, the approximation holds true. We refer to a number of repeated swaps between the system and a sequence of identical virtual qubits of the machine as a \emph{stage}. In this case, the exchange leads to a population increase of $\Delta p_n^{(i)} := p^{(i+1,i)}\subtiny{0}{0}{CH} - p^{(i)}\subtiny{0}{0}{S}$ in the energy level $E\subtiny{0}{0}{S}^{(i)}$. This, in turn, allows us to proceed without tracking the full state evolution of the machines when calculating the energy costs associated with both hot and cold machines at each stage $n$, which only depends upon the initial and final states of the target system rather than the more complicated machine states.

As a result, the energy change of each system and within each subspace at stage $n$ of the cooling procedure is thus
\begin{align}
    \Delta E\suptiny{0}{0}{(i)}\subtiny{0}{0}{S_n}= -\omega\subtiny{0}{0}{S}\suptiny{0}{0}{(i)} \Delta p\suptiny{0}{0}{(i)}_n, \quad\quad\quad \Delta E\subtiny{0}{0}{C_n}= \omega\subtiny{0}{0}{C_n}\suptiny{0}{0}{(i)} \Delta p\suptiny{0}{0}{(i)}_n, \quad\quad\quad \Delta E\suptiny{0}{0}{(i)}_{H_n}= -\omega\subtiny{0}{0}{H_n}\suptiny{0}{0}{(i)} \Delta p\suptiny{0}{0}{(i)}_n,
    \label{eq:energy exchange Inc subspace i}
\end{align}
where $\omega\subtiny{0}{0}{X}\suptiny{0}{0}{(i)} := E\subtiny{0}{0}{X}\suptiny{1}{0}{(i\!+\!1)}- E\subtiny{0}{0}{X}\suptiny{1}{0}{(i)}$ with $\{E\subtiny{0}{0}{X}\suptiny{1}{0}{(i)}\}_{i=0}^{d-1}$ are the energy eigenvalues of system $X$ sorted in non-decreasing order, and $\Delta p\suptiny{0}{0}{(i)}_n$ is the population exchange between energy levels $i$ and $i+1$ of the system at stage $n$\@. Combining Eqs.~\eqref{eq:qudit hamiltonian} and~\eqref{eq:energy exchange Inc subspace i} yields
\begin{equation}
\Delta E\suptiny{0}{0}{(i)}\subtiny{0}{0}{C_n}=\left( \left( \gamma+1 \right) \lambda_n -\,\gamma \right) \Delta E\suptiny{0}{0}{(i)}\subtiny{0}{0}{S_n} \quad\quad \text{and} \quad\quad \Delta E\suptiny{0}{0}{(i)}\subtiny{0}{0}{H_n}= \left(\gamma+1\right) \left(\lambda_n -1\right) \Delta E\suptiny{0}{0}{(i)}\subtiny{0}{0}{S_n}. 
 \label{eq:energy exchange subspace i qudit }
\end{equation}  
Now, using the fact that the total energy cost associated with each system can be calculated by summing the energy changes in each subspace, i.e., $\Delta E\subtiny{0}{0}{X_n}=\, \sum_{i=0}^{d-1}\, \Delta E\suptiny{0}{0}{(i)}\subtiny{0}{0}{X_n}$, at each step, it follows (from energy conservation) that one can calculate both the energy drawn from the hot bath and the heat dissipated into the cold bath for any such cooling procedure in terms of the energy change of the target system itself, i.e., $\Delta E\subtiny{0}{0}{C_n}=\left( \left( \gamma+1 \right) \lambda_n -\,\gamma \right) \Delta E\subtiny{0}{0}{S_n}$ and $\Delta E\subtiny{0}{0}{H_n}= \left(\gamma+1\right) \left(\lambda_n -1\right) \Delta E\subtiny{0}{0}{S_n}$\@. 

Thus, we need to determine the final state of the target system after each stage $n$\@. Based on the type of interaction specified in Eq.~\eqref{eq::incoherentquditinteraction}, we can show that this final state depends on the \emph{virtual Gibbs ratios} of the virtual machine qubits that non-trivially interact with the target system. The virtual Gibbs ratio associated with subspace $i$ at stage $n$ is given by $g\suptiny{0}{0}{(i)}_n := \frac{p^{(i+1,i)}_{C_n H_n}}{p^{(i,i+1)}_{C_n H_n}}=e^{-\beta \lambda_n \omega\subtiny{0}{0}{S}\suptiny{0}{0}{(i)}}$, where for instance $p^{(i+1,i)}_{C_n H_n}$ denotes the population in the subspace associated with $\ket{i+1, i}\subtiny{0}{0}{C_n H_n}$\@. As the number of repeated interactions between the $i^{\text{th}}$ subspace of the target and that of the machine increases, the virtual Gibbs ratio associated with the target subspace $i$ at stage $n$ will converge accordingly, i.e., $\frac{p^{(i\!+\!1)}_{S_n}}{p^{(i)}_{S_n}} \to e^{-\beta \lambda_n \omega\subtiny{0}{0}{S}\suptiny{0}{0}{(i)}}$\@. It is finally straightforward to see that the thermal state $\tau(\beta,\lambda_n H\subtiny{0}{0}{S})$ satisfies this condition, namely that the Gibbs ratio between each pair of neighbouring energy levels in each stage is given by $e^{-\beta \lambda_n \omega\subtiny{0}{0}{S}\suptiny{0}{0}{(i)}}$\@. It is worth mentioning here that in this protocol, rather than the individual states of both hot and cold machines per se, it is the virtual qubits of the total machine that play the most significant role in the reachable states of the target system and the corresponding energy cost.

For the sake of notational simplicity, we will denote thermal states as $\tau_n(\beta):= \tau(\beta, \lambda_{n} H\subtiny{0}{0}{S})$, with the boundary points $\tau\subtiny{0}{0}N(\beta):= \tau(\beta, \lambda_{\mathrm{f}} H\subtiny{0}{0}{S})$ and $\tau\subtiny{0}{0}0(\beta):= \tau(\beta, H\subtiny{0}{0}{S})$\@. With this, the energy dissipated into the cold machine can be calculated as 
\begin{align}
\Delta E\subtiny{0}{0}{C} &= -\sum_{n=1}^N\, \Delta E\subtiny{0}{0}{C_n}= -\sum_{n=1}^N\, \left( \left( \gamma+1 \right) \lambda_n -\,\gamma \right)\Delta E\subtiny{0}{0}{S_n}\, \nonumber\\*
&=-\sum_{n=1}^N\,  \left( \left( \gamma+1 \right) \lambda_n -\,\gamma \right)\tr{H\subtiny{0}{0}{S} \left\{ \tau_n(\beta)-\tau_{n-1}(\beta)\right\} } \nonumber\\*
    &=-\,\left(\gamma+1\right) \sum_{n=1}^N\,\tr{\lambda_n \,H\subtiny{0}{0}{S} \left\{\tau_n(\beta)-\tau_{n-1}(\beta)\right\}}
     +\,\gamma \,\tr{ H\subtiny{0}{0}{S} \left\{\tau_{\mathrm{f}}(\beta )-\tau\subtiny{0}{0}0(\beta)\right\}}\nonumber\\*
      &=\,\frac{\left(\gamma+1\right)}{\beta} \sum_{n=1}^N\,\bigg\{ D \left(\tau_{n-1}(\beta)\nr\|\nr\tau_{n}(\beta)\right)- S\left(\tau_{n}(\beta)\right)+S\left(\tau_{n-1}(\beta)\right)\bigg\}+\,\gamma \,\tr{H\subtiny{0}{0}{S} \left\{\tau_{\mathrm{f}}(\beta)-\tau\subtiny{0}{0}0(\beta)\right\}}\nonumber\\*
      &=\,\frac{\left(\gamma+1\right)}{\beta} \sum_{n=1}^N\,D \left(\tau_{n-1}(\beta)\nr\|\nr\tau_{n}(\beta)\right)
      -\frac{\left(\gamma+1\right)}{\beta} \Delta S\subtiny{0}{0}{S} +\, \gamma\, \Delta E\subtiny{0}{0}{S},
     \label{eq: energy exch C inco}
\end{align}
where we have made use of $D \left(\varrho\nr\|\nr\tau(\beta)\right)= -\beta\left[E(\tau(\beta))-E(\varrho)\right] + S(\tau(\beta))-S(\varrho)$ in the third line. 

Similarly, the energy drawn from the hot bath can be calculated as
\begin{align}
  \Delta E\subtiny{0}{0}{H}&= \sum_{n=1}^N\, \tr{H\subtiny{0}{0}{H_n} \left(\tau_n(\beta)-\tau_{n-1}(\beta)\right)}\, \nonumber\\*
  &=\,\left(\gamma+1\right)\sum_{n=1}^N\, \tr{(\lambda_n -1)\, H\subtiny{0}{0}{S} \left\{\tau_n(\beta)-\tau_{n-1}(\beta)\right\}}\, \nonumber\\*
    & =\,\left(\gamma+1\right) \sum_{n=1}^N\,\bigg\{ \tr{\lambda_n \,H\subtiny{0}{0}{S} \left\{\tau_n(\beta)-\tau_{n-1}(\beta)\right\}} \bigg\}
  -\,\left(\gamma+1\right) \,\tr{H\subtiny{0}{0}{S} \left\{\tau_{\mathrm{f}}(\beta)-\tau\subtiny{0}{0}0(\beta)\right\}}\nonumber\\*
    &=\,-\frac{\gamma+1}{\beta} \sum_{n=1}^N\,\bigg\{D \left(\tau_{n-1}(\beta)\nr\|\nr\tau_{n}(\beta)\right)+ S\left(\tau_{n-1}(\beta)\right)-S\left(\tau_{n}(\beta)\right)\bigg\}
   -\,\left(\gamma+1\right) \,\tr{H\subtiny{0}{0}{S} \left\{\tau_{\mathrm{f}}(\beta)-\tau\subtiny{0}{0}0(\beta)\right\}}\nonumber\\*
    &=\,-\frac{\gamma+1}{\beta} \sum_{n=1}^N\,\bigg\{D \left(\tau_{n-1}(\beta)\nr\|\nr\tau_{n}(\beta)\right)+ S\left(\tau_{n}(\beta)\right)-S\left(\tau_{n-1}(\beta)\right)\bigg\}
   -\,\left(\gamma+1\right) \,\tr{H\subtiny{0}{0}{S} \left\{\tau_{\mathrm{f}}(\beta)-\tau\subtiny{0}{0}0(\beta)\right\}}\nonumber\\*
      &=\,-\frac{\gamma+1}{\beta} \sum_{n=1}^N\, \bigg\{D(\tau_{n-1}(\beta)\nr\|\nr\tau_{n}(\beta))\bigg\}
  -\,\left(\gamma+1\right)\left[ \,\tr{H\subtiny{0}{0}{S} \left\{\tau_{\mathrm{f}}(\beta)-\tau\subtiny{0}{0}{0}(\beta)\right\}}-\frac{1}{\beta}\left\{S(\tau_{\mathrm{f}}(\beta))-S(\tau\subtiny{0}{0}{0}(\beta))\right\}\right].
     \label{eq: energy exch H inco1}    
\end{align}
Using the definition of free energy, $F\subtiny{0}{0}{X}^\beta(\varrho):= E\subtiny{0}{0}{X}(\varrho)- \beta^{-1} S(\varrho)$, Eq.~(\ref{eq: energy exch H inco1}) can be written as
\begin{align}
\Delta F\subtiny{0}{0}{S}^\beta \,+\eta \,\Delta E\subtiny{0}{0}{H}&=\,-\beta^{-1} \sum_{n=1}^N\, D\left(\tau_{n-1}(\beta)\nr\|\nr\tau_{n}(\beta)\right),
     \label{eq: energy exch H inco2}    
\end{align}
where $\eta = (\gamma +1)^{-1}$ is the Carnot efficiency.

Using the same methods as in SM Sec.~\ref{app::thermodynamiclengthcoherent}, the relative entropy terms in both Eqs.~\eqref{eq: energy exch C inco} and~\eqref{eq: energy exch H inco1} can be bounded in terms of the thermodynamic length, leading to the expressions stated in Eqs.~\eqref{eq: 3q incoh energy cost C} and~\eqref{eq::appthmincoherenthot}, respectively, as required. 

This provides a protocol for cooling a qudit target with qudit hot and cold machines, for which the energy cost can be calculated exactly. For the specific machine Hamiltonians and interactions considered [see Eqs.~\eqref{eq:qudit hamiltonian} and~\eqref{eq::incoherentquditinteraction}], the thermodynamic length expression is optimal in terms of heat dissipation; this therefore completes the first part of the proof. However, global optimality here is not guaranteed since the family of unitaries from which the protocol is derived only corresponds to a subset of energy-conserving ones. Nonetheless, in the qubit case, the analysis simplifies and the class of unitaries considered indeed covers the full orbit of energy-conserving ones (that do not create coherences, which could only be detrimental). Thus, to complete the proof, we can now finally argue for the general optimality of said cooling procedure when all systems are qubits.\\

\textbf{Three qubit case.---}Recall that in the coherent-control setting, the state of lowest achievable temperature is obtained by swapping each qubit system with successively colder machine qubits. However, such swaps are prohibited in the incoherent setting as they are not energy conserving. Nonetheless, as mentioned above, here one can swap the system with \emph{virtual qubits} of the hot-and-cold machines. Since any such subspace has norm strictly less than one, each such swap will lead to the system qubit being at strictly higher temperature than that of the virtual machine qubit. However, in the limit of infinitely many repetitions, the temperature of the system-qubit subspace of interest converges to the virtual temperature of the machine-qubit subspace~\cite{Clivaz_2019L,Clivaz_2019E}. As we are interested in finite resources, we assume a finite but sufficiently large number of swaps such that the error is acceptable. 

With this in mind, following a similar logic to that presented regarding the coherent-control setting, we assume that the local energy gaps of all qubit systems during interaction stage $n$ are given by 
\begin{equation}\label{eq::energygapincoherentqubits}
   E\subtiny{0}{0}{S}= E, \;\; E_{C_n}=\lambda_n \, E \,\quad \text{and} \quad\,E_{H_n}= \, (\lambda_n-1) \,E, 
\end{equation}    
where $E \in \mathbb{R}^+$\@. In order to make it possible for the target system to be cooled, the condition $E_{C_n}\, >\,E\subtiny{0}{0}{S}$ must be satisfied~\cite{Clivaz_2019E}. In this case, at each stage $n$, the degenerate subspace is spanned by the vectors $\{\ket{010},\,\ket{101}\}$, where we write $\ket{ijk}:=\, \ket{i}\subtiny{0}{0}{S}\otimes \ket{j}\subtiny{0}{0}{C}\otimes\ket{k}\subtiny{0}{0}{H}$\@. 

We now seek the types of energy-conserving unitaries that can cool the target system given this structure. We assume that the cooling scheme is Markovian in the sense that it can be represented as a collision model with the hot and cold baths being completely reset after each step of controlled evolution. As such, any possible correlations generated between the target and machines cannot be used in the next step and are therefore irrelevant. Moreover, creating correlations and coherences from initially uncorrelated thermal states incurs an energy cost~\cite{Huber_2015,BruschiPerarnauLlobetFriisHovhannisyanHuber2015, Misra_2016, Bakhshinezhad_2019}. Since we are seeking the minimum such cost, it follows that we need only consider energy-conserving unitaries that permute the populations of the eigenstates. All possible such unitaries can be generated by Hamiltonians of the following form
\begin{align}
\widetilde{H}_{\textup{int}} =\ketbra{010}{101}+\ketbra{101}{010},
\label{eq:cooling interaction}
\end{align}
since acting with $\widetilde{U}(t) =\, \mathrm{exp}[-i \widetilde{H}_{\textup{int}}\, t]$ does not generate any coherence in the marginals of the output state. It is worth mentioning that cooling in the incoherent-control setting has been investigated completely for different energy-level structures and all energy-conserving unitaries in Ref.~\cite{Clivaz_2019E}. There it was shown that the above energy-level structure and interaction Hamiltonian are the \emph{only} ones that can cool the target system. In other words, as far as cooling with incoherent control for qubits is concerned it is in general sufficient to consider the combination of machine structure given by Eq.~\eqref{eq::energygapincoherentqubits} and interactions in Eq.~\eqref{eq:cooling interaction}. With this in mind we now prove optimality. 

Due to the energy-conserving nature of the unitary interaction, if the population of the ground state of the target system is increased by $\Delta p_n$, all local energy changes at stage $n$ can be calculated as 
\begin{align}
    \Delta E_{S_n}= -E\subtiny{0}{0}{S} \Delta p_n, ~~ \Delta E_{C_n}= E_{C_n} \Delta p_n, ~~\Delta E_{H_n}= -E_{H_n} \Delta p_n,
    \label{eq:energy exchange Inc}
\end{align}
where the energy change of subsystem $X$ is defined by $\Delta E\subtiny{0}{0}{X}=\, \tr{H\subtiny{-1}{0}{X}\, (\varrho^\prime\subtiny{0}{0}{X}-\,\varrho\subtiny{0}{0}{X})}$\@. Thus, the energy cost per unit of population exchange only depends upon the energy-level structure; this is a special feature that holds for qubits only and does not extend to higher-dimensional systems and hence represents a roadblock for generalisation. In order to minimise the energy cost of cooling, one should therefore use the smallest available energy gap to cool the target system as much as possible at each stage.

Note that, since such an energy-conserving unitary only acts non-trivially in a (strict) subspace, whose population cannot be unity, its effective operation on $S$ and $CH$ considered as a whole is a partial swap (cf. the full swap with the machine in the coherent-control case). Nonetheless, if such a partial swap is repeated a sufficiently many times, refreshing the hot and cold machines each time, the state of the system converges to that which would be obtained by completely swapping the target with a \emph{virtual} qubit of the machine, i.e., spanned by the vectors $\{ \ket{01}\subtiny{0}{0}{CH} , \ket{10}\subtiny{0}{0}{CH}\}$~\cite{Clivaz_2019L,Clivaz_2019E}. In order to cool the target system to inverse temperature $\beta_n := \frac{E_n}{E\subtiny{0}{0}{S}}\beta$ at stage $n$ in the incoherent-control setting, the energy gaps of the cold and hot machines must satisfy $E_{C_n} \geq (\gamma+1) E_n -\gamma E\subtiny{0}{0}{S}$ and $E_{H_n} \geq (\gamma+1) (E_n -\,E\subtiny{0}{0}{S})$ , where $\gamma:=\frac{\beta\subtiny{0}{0}{H}}{\beta-\beta\subtiny{0}{0}{H}} (=\eta^{-1}-1)$ (see Appendix F 2 in Ref.~\cite{Taranto_2023}). Since the energy cost per population exchange depends linearly upon the energy gaps themselves, it is clear that one must use the smallest gaps possible to cool the system at each step in order to minimise the energy cost. Thus, the optimal machine energy structure at each stage $n$ is given by $E_{C_n} = (\gamma+1) E_n -\gamma E\subtiny{0}{0}{S}$ and $E_{H_n}=(\gamma+1) (E_n -\,E\subtiny{0}{0}{S})$.

We now move to focus on optimality of the transformation itself [generated by Hamiltonians of the form given in Eq.~\eqref{eq:cooling interaction}]. The diagonal energy-conserving transformations considered here for the qubit case can generally be divided into those that increase the ground-state population of the target and those that decrease it, which respectively correspond to \emph{cooling} and \emph{heating} the target system. We will now show that heating up the system at any time cannot help to reduce the energy cost of the cooling process. And since the overall protocol proceeds in a Markovian fashion, this therefore completes the proof.

We proceed by way of contradiction. Consider a qubit target system interacting with fresh qubit hot and cold thermal machines at each time. We assume that each machine can be a qubit system whose energy gap is finite, but we have access to infinitely many copies of them. We claim here that the optimal cooling procedure is to cool the target system at each step. Suppose, for the sake of contradiction, that this is not true and that heating the target at some point can help to reduce the energy cost in reaching a desired state at the end of the protocol.

We begin by assuming that the virtual qubits of the machines are ordered in non-decreasing fashion and that we use the smallest possible gap that permits cooling the target from $\frac{p\suptiny{0}{0}{(n-1)}_1}{p\suptiny{0}{0}{(n-1)}_0}= e^{-\beta E_{n-1} }$ to  $\frac{p\suptiny{0}{0}{(n)}_1}{p\suptiny{0}{0}{(n)}_0}= e^{-\beta E_n }$ in stage $n$\@. As mentioned above, the system must interact here with a machine whose cold subsystem has an energy gap of at least $E\subtiny{0}{0}{C_n}=\, \left (\gamma+1\right) E_n -\gamma E\subtiny{0}{0}{S}$\@. The energy dissipated into the cold bath of this stage is simply the product of the cold-machine energy gap and the population transfer, which leads to $\Delta E\subtiny{0}{0}{C_n}/\Delta p_n\geq E\subtiny{0}{0}{C_n}$ and $\Delta E\subtiny{0}{0}{C_n} \geq 0$\@. In the next stage $n+1$, we will investigate all possible ways of heating up the target via energy-conserving unitaries. Since heating the target is equivalent to cooling the cold machine, it is, \emph{a priori} possible that such a strategy could lead to a reduced energy cost in the long run.

However, this is not the case, as we now show. In order to cool down the cold machine, the target system must interact with a virtual qubit whose population ratio exceeds $e^{-\beta E_{n}}$\@. For any such heating of the target to reduce the energy cost, one must show that there exists a choice of energy gaps $E\subtiny{0}{0}{C_{n+1}}$ (and corresponding $E\subtiny{0}{0}{H_{n+1}}$) where $E\subtiny{0}{0}{C_{n+1}}\geq\, E\subtiny{0}{0}{C_{n}}$ and $\Delta E\subtiny{0}{0}{C_{n+1}}/\Delta p_{n+1}= E\subtiny{0}{0}{C_{n+1}}$ such that $\Delta E\subtiny{0}{0}{C_{n+1} }\leq 0$\@. Intuitively, these constraints imply that the energy cost per population exchange in the cooling stage is lower than that of the heating one. If this were true, then one could extract energy from the cold system via a cyclic process, which contradicts the second law. In the following, we will formally show that it is impossible to find such a process by considering all possible energy-level structures and energy-conserving unitaries on three qubits. We finally conclude that heating the system at any point can only serve to increase the energy cost of the overall protocol, and thus the optimal overall procedure can only be a concatenation of cooling steps. 

We proceed on a case-by-case basis.
\begin{enumerate}[a)]
\item $E\subtiny{0}{0}{X}= E\subtiny{0}{0}{Y}$ where $X,Y \in\, \{S,\,C,\,H\}$: In this case, the degenerate subspace is spanned by $\{\ket{001},\, \ket{010}, \, \ket{100}\}$\@. Since $\beta\subtiny{0}{0}{H} \leq \beta$, one can reduce the energy dissipated by the cold machine via a bipartite interaction with the target system at inverse temperature $\beta_n= \frac{E_n}{E\subtiny{0}{0}{S}}\beta\geq \beta$\@. Here, the energy transferred to the cold machine per population exchange is determined by $E\subtiny{0}{0}{S} < E\subtiny{0}{0}{C_n}$\@. Thus, the energy cost cannot be reduced.
\item $E\subtiny{0}{0}{C}=\, E\subtiny{0}{0}{S}+\, E\subtiny{0}{0}{H}$: In this case, the degenerate subspace is spanned by $\{\ket{101},\, \ket{010}\}$\@. From Ref.~\cite{Taranto_2023}, in order to heat up the target system, the energy gap of the cold machine must be $E\subtiny{0}{0}{C_{n+1}}\leq\, \left (\gamma+1\right) E_n -\,\gamma\, E\subtiny{0}{0}{S}=\,E\subtiny{0}{0}{C_{n}}$\@. Thus, the energy cost cannot be reduced. 
\item $E\subtiny{0}{0}{C}=\, 2E\subtiny{0}{0}{S}=\, 2E\subtiny{0}{0}{H}$: In this case, there are three different degenerate subspaces. One of them discussed in case b); the other two degenerate subspaces are spanned by $\{\ket{001},\, \ket{100}\}$ and $\{\ket{011},\, \ket{110}\}$, respectively. In either case, the system can be heated without interacting with the cold machine (i.e., with only a bipartite interaction with the hot machine), which does not affect the energy dissipated by the cold machine.  
\item $E\subtiny{0}{0}{H}=\, E\subtiny{0}{0}{S}+\, E\subtiny{0}{0}{C}$: In this case, the degenerate subspace is spanned by $\{\ket{110},\, \ket{001}\}$\@. Here, if the target is heated, so too is the cold machine. Thus, this setting also cannot help to reduce the energy cost.
\end{enumerate}
We can now conclude that in 3-qubit incoherent-control cooling scenarios, heating up the target system at any point cannot reduce the total energy cost of cooling, and so the optimal process must be a concatenation of cooling steps. \end{proof}


\section{Role of correlations in the incoherent-control paradigm}\label{app::incoherentcorrelations}

\begin{proof} \textbf{(Theorem~\ref{thm::incoherentcorrelations}).} We start from the strong subadditivity of the von Neumann entropy~\cite{LiebRuskai1973},
\begin{align}
    S\subtiny{0}{0}{SCH} + S\subtiny{0}{0}{S} \leq S\subtiny{0}{0}{SC} + S\subtiny{0}{0}{SH},
    \label{eq:SSA app}
\end{align}
where we use the notation $S\subtiny{0}{0}{X} := S(\varrho\subtiny{0}{0}{X})$. Since we begin with an initially uncorrelated global state, the inequality evaluated on $\varrho\subtiny{0}{0}{SCH} = \tau\subtiny{0}{0}{S}(\beta)\otimes\tau\subtiny{0}{0}{C}(\beta)\otimes\tau\subtiny{0}{0}{H}(\beta\subtiny{0}{0}{H})$ reduces to equality, i.e., $ S\subtiny{0}{0}{SCH}\suptiny{0}{0}{0}  + S\subtiny{0}{0}{S}\suptiny{0}{0}{0} \,= S\subtiny{0}{0}{SC}\suptiny{0}{0}{0}  + S\subtiny{0}{0}{SH}\suptiny{0}{0}{0} $. In addition, since the global evolution is unitary, we have that $\Delta S\subtiny{0}{0}{SCH} = 0$, where $\Delta S\subtiny{0}{0}{X} := S\subtiny{0}{0}{X}^\prime -\, S\subtiny{0}{0}{X}\suptiny{0}{0}{0}$. Hence
\begin{align}
    \Delta S\subtiny{0}{0}{S} \leq \Delta S\subtiny{0}{0}{SC} + \Delta S\subtiny{0}{0}{SH},
\end{align}  
Writing $I\subtiny{0}{0}{XY} := I(X\!:\!Y)_{\varrho\subtiny{0}{0}{XY}} =S(\varrho\subtiny{0}{0}{X})+S(\varrho\subtiny{0}{0}{Y})-S(\varrho\subtiny{0}{0}{XY})$ then yields
\begin{align}\label{eq::deltasinequality-1}
    - \Delta S\subtiny{0}{0}{S} \leq - \Delta I\subtiny{0}{0}{SC} - \Delta I\subtiny{0}{0}{SH} + \Delta S\subtiny{0}{0}{C} + \Delta S\subtiny{0}{0}{H}.
\end{align}
By symmetry, the same arguments as above can be used to derive
\begin{align}
    - \Delta S\subtiny{0}{0}{C} \leq - \Delta I\subtiny{0}{0}{SC} - \Delta I\subtiny{0}{0}{CH} + \Delta S\subtiny{0}{0}{S} + \Delta S\subtiny{0}{0}{H}, \label{eq::deltasinequality-2}\\[1mm]
    - \Delta S\subtiny{0}{0}{H} \leq - \Delta I\subtiny{0}{0}{SH} - \Delta I\subtiny{0}{0}{CH} + \Delta S\subtiny{0}{0}{S} + \Delta S\subtiny{0}{0}{C}. \label{eq::deltasinequality-3}
\end{align}
Combining Eqs.~\eqref{eq::deltasinequality-1},~\eqref{eq::deltasinequality-2}, and~\eqref{eq::deltasinequality-3} then leads to
\begin{align}\label{eq::deltasinequality-6}
    \sum\subtiny{0}{0}{X\!\in\!\{\!S\!,\!C\!,\!H\!\}} \Delta S\subtiny{0}{0}{X} \geq \frac{2}{3} \sum_\alpha \Delta I_\alpha,
\end{align}
where $\alpha \in \{ SC, SH, CH\}$. Finally, recall the free-energy difference $\Delta F\subtiny{0}{0}{X}^{\beta} := \Delta E\subtiny{0}{0}{X} - \beta^{-1} \Delta S\subtiny{0}{0}{X}$. If we consider the total energy change of the entire system as work done on the total system by an external agent, i.e., $W := -\sum\subtiny{0}{0}{X} \Delta E\subtiny{0}{0}{X}$, then Eq.~\eqref{eq::deltasinequality-6} can be written in the form
\begin{align}\label{eq::deltasinequality-7}
    \sum\subtiny{0}{0}{X\!\in\!\{\!S\!,\!C\!,\!H\!\}} \Delta F\subtiny{0}{0}{X}^{\beta} \leq - W - \frac{2}{3} \beta^{-1}\sum_\alpha \Delta I_\alpha.
\end{align}
Since the total energy is conserved, we have $W = 0$\@. Substituting this into Eq.~\eqref{eq::deltasinequality-7} asserts the claim. \end{proof}

\end{document}